\documentclass[runningheads]{llncs}

\usepackage{graphicx}

\usepackage{booktabs}
\usepackage{float}
\usepackage{threeparttable}
\usepackage{xspace}
\usepackage{adjustbox}
\usepackage{subcaption}
\usepackage{relsize}
\usepackage{svg}
\usepackage{array}
\usepackage{amsfonts}
\usepackage{tabularx}
\usepackage[inline]{enumitem}

\usepackage{pdflscape}
\usepackage{comment}
\usepackage{multirow}

\usepackage{glossaries}
\setacronymstyle{long-short}
\glsdisablehyper
\newacronym{gan}{GAN}{Generative Adversarial Network}
\newacronym{cgan}{CGAN}{Conditional GAN}
\newacronym{irgan}{IRGAN}{Information Retrieval GAN}
\newacronym{cfgan}{CFGAN}{Collaborative Filtering GAN}
\newacronym{sgd}{SGD}{Stochastic Gradient Descent}
\newacronym{zr}{ZR}{Zero Reconstruction}
\newacronym{pm}{PM}{Partial Masking}
\newacronym{zp}{ZP}{Zero Reconstruction and Partial Masking}
\newacronym{ott}{OTT}{Over-The-Top Media service}
\newacronym{vod}{VoD}{Video-on-Demand}
\newacronym{utc}{UTC}{Coordinated Universal Time}
\newacronym{vcpu}{vCPU}{virtual Central Processing Unit}
\newacronym[
longplural={User-Rating Matrices}
]
{urm}{URM}{User-Rating Matrix}

\newacronym{map}{MAP}{Mean Average Precision}
\newacronym{roc}{ROC}{Receiver Operating Characteristic curve}
\newacronym{rocauc}{ROC-AUC}{Receiver Operating Characteristic Area Under the Curve}
\newacronym{ndcg}{NDCG}{Normalized Discounted Cumulative Gain}
\newacronym{mrr}{MRR}{Mean Reciprocal Rank}
\newacronym{rs}{RS}{Recommender System}
\newacronym{ctr}{CTR}{Click Through Rate}

\usepackage{hyperref}

\usepackage[misc]{ifsym}

\makeatletter
\renewcommand{\Itemautorefname}{\@gobble}
\makeatother
\newcommand{\x}{%
 \ensuremath{\boldsymbol{x}}\xspace
}
\newcommand{\pdata}{%
 \ensuremath{p_{data}(\x)}\xspace
}

\newcommand{\z}{%
 \ensuremath{\boldsymbol{z}}\xspace
}
\newcommand*{\pz}{%
 \ensuremath{p_{z}(\z)}\xspace
}

\newcommand{\pg}{%
 \ensuremath{p_{g}(\x)}\xspace
}
\newcommand{\gen}{%
 \ensuremath{G}\xspace
}
\newcommand{\genparam}{%
 \ensuremath{\theta_{g}}\xspace 
}
\newcommand{\genfunc}{%
 \ensuremath{\gen(\z, \genparam)}\xspace 
}

\newcommand{\disc}{%
 \ensuremath{D}\xspace
}
\newcommand{\discparam}{%
 \ensuremath{\theta_{d}}\xspace
}
\newcommand{\discfunc}{%
 \ensuremath{\disc(\x,\discparam)}\xspace
}

\newcommand{\ganloss}{%
 \ensuremath{l_{GAN}}\xspace
}

\newcommand{\cond}{%
 \ensuremath{\boldsymbol{c}}\xspace
}

\newcommand{\replications}{$30$\xspace}

\newcommand{\idest}{i.e.,\xspace}
\newcommand{\eg}{e.g.,\xspace}

\newcommand{\rqone}{\textbf{RQ1}\xspace}
\newcommand{\rqtwo}{\textbf{RQ2}\xspace}
\newcommand{\rqthree}{\textbf{RQ3}\xspace}

\begin{document}
\title{An Evaluation Study of Generative Adversarial Networks for Collaborative Filtering}
\titlerunning{An Evaluation of GANs for Collaborative Filtering}
%
\author{%
    Fernando B. {P\'erez Maurera}\inst{1,2}\textsuperscript{,\Letter}\orcidID{0000-0001-6578-7404} 
    \and
    Maurizio {Ferrari Dacrema}\inst{1}\orcidID{0000-0001-7103-2788} 
    \and
    Paolo Cremonesi\inst{1}\orcidID{0000-0002-1253-8081}
}

\authorrunning{F. B. {P\'erez Maurera} et al.}
%
\institute{%
    Politecnico di Milano, Milan, Italy \\
    \email{\{fernandobenjamin.perez,maurizio.ferrari,paolo.cremonesi\}@polimi.it}
    \and
    ContentWise, Milan, Italy \\%
    \email{fernando.perez@contentwise.com}
}
\maketitle              
%
%
%
\begin{abstract}

This work explores the reproducibility of CFGAN. 
CFGAN and its family of models (TagRec, MTPR, and CRGAN) learn to generate personalized and fake-but-realistic rankings of preferences for top-N recommendations by using previous interactions. 
This work successfully replicates the results published in the original paper and discusses the impact of certain differences between the CFGAN framework and the model used in the original evaluation. The absence of random noise and the use of real user profiles as condition vectors leaves the generator prone to learn a degenerate solution in which the output vector is identical to the input vector, therefore, behaving essentially as a simple autoencoder. 
The work further expands the experimental analysis comparing CFGAN against a selection of simple and well-known properly optimized baselines, observing that CFGAN is not consistently competitive against them despite its high computational cost. 
To ensure the reproducibility of these analyses, this work describes the experimental methodology and publishes all datasets and source code. 

\keywords{%
    Generative Adversarial Networks
    \and Recommender Systems
    \and Collaborative Filtering
    \and Reproducibility
}
\end{abstract}
\section{Introduction}
\label{sec:introduction}
In recent years, \glspl{gan} have become the state-of-the-art technique inside the group of generative methods, \idest methods that learn how to generate fake data from the real one. Their primary use has been in the computer vision domain \cite{gan,image-to-image-conditional-gan,stylegan,stylegan2}. 
They have also been used in Information Retrieval \cite{irgan} and Recommender Systems, the most notable example being \gls{cfgan} \cite{cfgan}, and the family of models based on it, such as TagRec \cite{tagrec-gan}, CRGAN \cite{crgan-adversarial-preference-learning-with-pairwise-comparisons}, MTPR \cite{mtpr-gan}, and \cite{cfgan-service-recommendations}.

This work contributes to the trend of evaluation studies in Machine Learning, Information Retrieval, and Recommender Systems domains \cite{a-troubling-analysis-of-reproducbility-and-progress-in-recsys-research,are-we-really-making-much-progress-a-worrying-analysis-of-recent-neural-recommendation-approaches,DBLP:journals/sigir/Lin18,DBLP:journals/sigir/Lin19,DBLP:conf/sigir/YangLYL19}. 
This work discusses the implications of certain differences between the \gls{cfgan} framework and the model that was used in the experimental evaluation, which would adversely affect its learning ability, providing a reference for future works. In particular, the generator is left prone to reach a degenerate solution and behave as a simple autoencoder, therefore, belonging to the same family of previous recommendation models such as \cite{slim,ease-r-embarrassingly-shallow-autoencoders-for-sparse-data}.
This discussion is based on the findings of \cite{troubling-trends-in-machine-learning-scholarship}, which highlights the importance of describing not only \emph{how} a model works, but also \emph{what} works and \emph{why} it works, as well as how experimental inquiries that aim to deepen our understanding are valuable research contributions even when no new algorithm is proposed. 
Furthermore, this work analyzes the replicability, reproducibility, and recommendation quality of \gls{cfgan} \cite{cfgan} as well as its numerical stability which is known to be a challenge for GANs \cite{seed-2-effects-of-random-seeds-on-the-accuracy-of-convolutional-neural-networks,seed-1-on-model-stability-as-a-function-of-random-seed}. 
The main research questions of this work are:
\begin{description}
    \item[\rqone:] Is \gls{cfgan} replicable and numerically stable? \idest does CFGAN achieve the claimed results using the same experimental setup as in \cite{cfgan}?
    \item[\rqtwo:] What is the impact of the differences between the \gls{cfgan} framework and the model used for the evaluation in \cite{cfgan}, and why do they raise theoretical and methodological concerns regarding the learning ability of the model?
    \item[\rqthree:] Is \gls{cfgan} reproducible, achieving the claimed recommendation quality when compared to properly-tuned baselines? How does \gls{cfgan} compare along other dimensions such as beyond-accuracy and scalability metrics?
\end{description}
\section{Collaborative Filtering Generative Adversarial Networks}
\label{sec:generative-adversarial-networks}

\glspl{gan} have been successfully applied to numerous prediction and classification tasks. This work addresses a family of generative models originated from \glspl{gan} used in Recommender Systems. Briefly, a \gls{gan}\footnote{The supplemental material\cite{supplemental_material} contains the formal formulation of \glspl{gan}.} consists of two neural networks that are trained together in an adversarial setting until they reach convergence. The first neural network is called the \emph{generator}, denoted as \gen, while the second network is called the \emph{discriminator}, denoted as \disc \cite{gan-evaluation,gans-an-overview,gan-communications-acm,gan}. 
\gls{cfgan}\footnote{For a detailed explanation of \gls{cfgan} we refer the reader to the reference article  \cite{cfgan}.} is the most notable \gls{gan} recommendation algorithm \cite{tagrec-gan,cfgan-service-recommendations}. Its main attribute is that it generates personalized user or item profiles, mainly  by solely using previous interactions, but is able to learn from sources of information as well \cite{cfgan}.

\subsubsection{\gls{cfgan} Training Process}
\label{subsec:generative-adversarial-networks:training-process} 
\begin{figure}[t]
    \centering
    \includegraphics[width=0.75\linewidth]{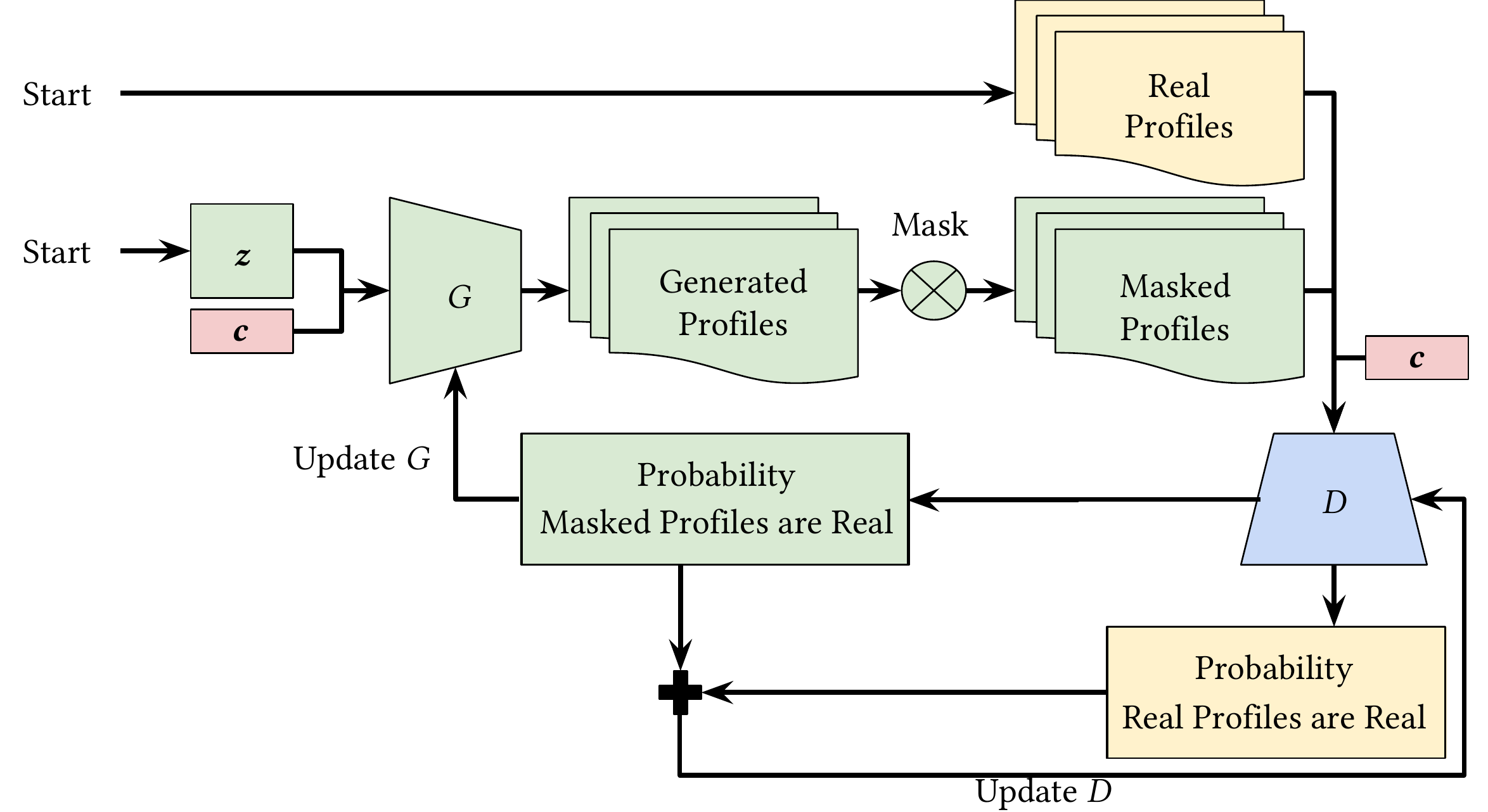}
    \caption{Training process of \gls{cfgan}. \gen, \disc, \z and \cond are the generator network, discriminator network, random noise, and condition vectors, respectively. Real profiles are not masked.}
    \label{fig:cfgan-training}
\end{figure}

\autoref{fig:cfgan-training} shows an illustration of the training process of \gls{cfgan}. Every epoch starts by feeding the generator \gen with random noise \z and a condition vector \cond. The generator creates preferences of users towards items (or vice versa) which are then masked (see Masking). The discriminator \disc then receives the real profiles, the masked profiles, and the condition. The discriminator tells the probability that each masked and real profiles come from the real data. The discriminator is updated based on how well it is able to correctly distinguish fake data from real data. The generator is updated based on how much it could generate \emph{fake but realistic} data.

\subsubsection{Modes} \gls{cfgan} has two modes: \emph{user-based} (\textit{u}) or \emph{item-based} (\textit{i}). The first learns to generate user profiles, while the second learns to generate item profiles.

\subsubsection{Masking} \gls{cfgan} applies a mask to the generated profiles by performing an element-wise product between these and the real profiles. If the variant is \emph{Partial Masking}, then the mask changes (see Variants).

\subsubsection{Architecture} Both the generator and discriminator of \gls{cfgan} are fully connected feed-forward neural networks independent from each other where each has its own hyper-parameters, \eg number of hidden layers, learning rate, regularization, and others. 
If the \emph{mode} is \emph{user-based}, then the number of input neurons is the number of \emph{items} in the dataset. Conversely, the number of input neurons for an \emph{item-based} \gls{cfgan} is the number of \emph{users} in the dataset.

\subsubsection{Recommendations} 

In a top-N item recommendation scenario, the trained generator creates user profiles containing the preference scores of users toward items. Recommendations are built by ranking the items from the highest to lowest score and selecting the top-N.

\subsubsection{Variants}
\label{subsubsec:generative-adversarial-networks:variants}
\gls{cfgan} has three variants:

\begin{itemize}
    \item \textit{\gls{zr}}: Changes the loss function of the generator. It ensures that a sample of non-interacted items are given zero-weights in the generated profiles.
    \item \textit{\gls{pm}}: The mask applied to the generated profiles combines the user profile and randomly-chosen unseen items.
    \item \textit{\gls{zp}}: Combines \gls{zr} and \gls{pm}.
\end{itemize}

\section{CFGAN Theoretical and Methodological Questions}
\label{sec:theoretical-and-methodological-issues}

This work highlights key differences between the initial description of \gls{cfgan} and the model used in the experimental evaluation of that same paper \cite{cfgan}. These differences were not discussed in the original paper but have significant implications on the model's ability to learn user or item preferences.

\subsection{Real Profiles as Condition Vectors} 

\subsubsection{What raises concerns?} In the experimental evaluation of \gls{cfgan}, the condition vector provided to both the generator and the discriminator is the real user/item profile, \idest the interactions that \gls{cfgan} is learning to generate. 

\subsubsection{Why is it a concern?} As a consequence, \gls{cfgan} is prone to generate a trivial solution. The generator could learn the identity function between the condition vector and the output, therefore easily deceiving the discriminator without learning to generate new profiles. On the other hand, the discriminator could learn that the generated user profile should be identical to the condition vector to be real, again learning a trivial function.
In practice, this will push the generator to behave as an \emph{autoencoder} \cite{DBLP:journals/ftml/KingmaW19/variational-autoencoder}, which reconstructs as output the same input (condition) it was provided with.

\subsubsection{How to avoid this concern?} Since the condition vector can contain any information, a simple strategy would be to use other feature data related to the items or users or other contextual information. In a pure collaborative recommendations scenario, where no features or context is available, a possible strategy is to change the condition vector to be the user/item classes (\idest unique identifiers) depending on the \gls{cfgan} mode. 
This decision is aligned with previous works on \glspl{gan} \cite{conditional-gan}. 
In Recommender Systems, using the user/item classes provides a mechanism to generate \emph{personalized recommendations} to every user. 
In contrast to the original \gls{cfgan} reference, using the user/item classes excludes the possibility that the generator and discriminator learn a trivial solution.

\subsection{No Random Noise} 

\subsubsection{What raises concerns?} The reference article states 
that the random noise is not provided as input to the generator in its experiments because the goal is to generate the single best recommendation list rather than multiple ones.

\subsubsection{Why is it a concern?} This violates the framework defined in the same article and the design principles of \glspl{gan}. In practice, discarding noise is problematic because it drastically reduces the input space and the generator will be trained on a very sparse set of user profiles. This assumes that the user profiles will not change, which will make \gls{cfgan} non-robust in a real application where the data change rapidly. This is known as the \emph{dataset shift} problem. Since the data change over time as new interactions are collected, and models are not continuously retrained, models should be robust to and be able to use the \emph{new} data that was not present during training \cite{dataset-shift-in-machine-learning,a-unifying-view-on-dataset-shift-in-classification}.

\subsubsection{How to avoid this concern?} Feed the generator with a random noise vector \z and the condition vector \cond. \z is drawn from a normal distribution with zero mean and unit variance, \idest $\z \sim \mathcal{N}(\mu,\,\sigma^{2})$ where $\mu = 0$ and $\sigma^{2} = 1$ as suggested by other works \cite{gan,conditional-gan}. The size of \z is a key element while training \glspl{gan}. However, previous works do not have consensus concerning the size of \z \cite{gans-an-overview}. We use a heuristic to set the size of the random vector and try different values depending on the number of input neurons:  50\%, 100\%, or 200\% of them. In practice, the condition \cond and the random vector \z are concatenated, and this new vector becomes the input to the first layer of the generator network.

\subsection{Methodological Questions} 

\subsubsection{What raises concerns?} The \gls{cfgan} description does not state how to choose the number of training epochs nor the stopping criterion for the training phase.

\subsubsection{Why is it a concern?} The number of training epochs and the stopping criterion are two key methodological aspects for most machine learning models. With the current \gls{gan} formulation, these two are defined by hand instead of automatically chosen by the continuous evaluation of \gls{gan}, which might lead to a non-optimal model, misuse of computational resources, and negatively affect the published results' replicability. There are well-known objective ways to measure the recommendation quality in offline scenarios without human intervention in the Recommender Systems domain, \eg with accuracy metrics. 

\subsubsection{How to avoid this concern?} Use an early-stopping mechanism based on the one used in previous works for other machine learning recommenders, such as matrix factorization or linear regression \cite{a-troubling-analysis-of-reproducbility-and-progress-in-recsys-research,are-we-really-making-much-progress-a-worrying-analysis-of-recent-neural-recommendation-approaches}. An early-stopping mechanism periodically evaluates \gls{cfgan} on validation data while \gls{cfgan} is being trained on train data. The training stops when the \gls{cfgan} quality does not improve over the best evaluation for a fixed number of evaluations. 

\section{Experimental Methodology}
\label{sec:experimental-methodology}

The experiments, results, and discussion are based on one of the following two experiments: \begin{enumerate*}[label=(\roman*)]
    \item execution of the source code provided in the \gls{cfgan} reference article as-is to assess the result replicability;
    \item hyper-parameter tuning of different recommenders using a well-known evaluation framework to study the reproducibility of the results and evaluate along different dimensions (see~\cite{a-troubling-analysis-of-reproducbility-and-progress-in-recsys-research,are-we-really-making-much-progress-a-worrying-analysis-of-recent-neural-recommendation-approaches})
\end{enumerate*}.
The source code of the experiments is available online~\footnote{\url{https://github.com/recsyspolimi/ecir-2022-an-evaluation-of-GAN-for-CF} and \cite{supplemental_material}.}.

\subsubsection{Datasets} The experiments use the same datasets\footnote{The Watcha \cite{cfgan} dataset was not provided with the reference article.} (a sub-sampled version of Ciao\footnote{The reference article does not provide instructions to reproduce this version of the dataset. We contacted the authors for clarifications but did not receive a reply.} \cite{cfgan,ciao-dataset}, ML100K \cite{movielens}, and ML1M \cite{movielens}) and splits (\textit{train} and \textit{test}) provided with the \gls{cfgan} reference article \cite{cfgan}. For scenarios that required a \emph{validation} split, we created one by applying the same strategy as the reference: random holdout at $80\%$ of the \emph{train} split. Given the modest size of these datasets, all experiments are done on the CPU.

\subsubsection{Technologies} The implementation of all experiments, is based on the evaluation framework published in \cite{a-troubling-analysis-of-reproducbility-and-progress-in-recsys-research}, which includes the implementation of some simple yet competitive state-of-the-art baselines for Recommender Systems. For the replication study, the original implementation has been used as provided. For the reproducibility study and the other experiments, the original \gls{cfgan} source code has been adapted to the framework with no changes to the core algorithm. 

\subsection{Methodology for the Replicability of CFGAN }
\label{subsec:experimental-methodology:original-code-execution}
The original \gls{cfgan} source code includes the implementation of \gls{cfgan} and its training loop using a fixed set of hyper-parameters that are dataset-dependent. 
The training procedure is the following: it fits a \gls{cfgan} recommender using the \emph{train} split of the selected dataset and evaluates the recommender using the \emph{test} split. 
With respect to the evaluation metrics, this source code evaluates \gls{cfgan} on \emph{accuracy} metrics: precision (PREC), recall (REC), \gls{mrr}, and \gls{ndcg} at recommendation list length $5$ and $20$. 
The limitations of this source code are the lack of the implementation of the baselines and the hyper-parameter tuning of all recommenders, \eg baselines and \gls{cfgan}. Due to this, the replication study is only possible for \gls{cfgan}. 

\subsection{Methodology for the Reproducibility of CFGAN}
\label{subsec:experimental-methodology:hyper-parameter-tuning}

The reproducibility study expands the original \gls{cfgan} evaluation by including: \begin{enumerate*}[label=(\roman*)]
    \item \label{item:reproducibility:baselines} new baselines that were shown to provide high recommendation quality;
    \item \label{item:reproducibility:hyper-parameter-tuning} a well-defined hyper-parameter optimization strategy;
    \item \label{item:reproducibility:early-stopping} a well-defined early-stopping strategy; and
    \item \label{item:reproducibility:metrics} a comparison against accuracy, beyond-accuracy, and scalability metrics
\end{enumerate*}.

In particular, the goal of \autoref{item:reproducibility:baselines} and \autoref{item:reproducibility:hyper-parameter-tuning} is to assess the recommendation quality of \gls{cfgan} against a wider set of recommendation models which are properly tuned under the same conditions. 
The models we report range from non-personalized, neighborhood-based, and non-neural machine learning approaches. This decision is aligned with results obtained by previous evaluation studies in the domain \cite{are-we-really-making-much-progress-a-worrying-analysis-of-recent-neural-recommendation-approaches,a-troubling-analysis-of-reproducbility-and-progress-in-recsys-research}.
Regarding the hyper-parameter optimization of \gls{cfgan}, it should be noted that the search-space described in the reference article, considering that it is done using a grid-search, contains more than $3 \cdot 10^8$ cases, which cannot be reproduced in a reasonable time. Due to this, this work adopts a different optimization strategy: Bayesian Search as used in \cite{a-troubling-analysis-of-reproducbility-and-progress-in-recsys-research}. The hyper-parameter ranges and distributions of \gls{cfgan} are reported in \autoref{tab:hyper-parameters}. The Bayesian Search starts with 16 initial random searches and performs a total of 50 cases for each algorithm. Each model in this search is fit with the \emph{train} split and evaluated against the \emph{validation} one. The best hyper-parameters are chosen as those with the highest \gls{ndcg} at 10. Once the optimal hyper-parameters set is chosen, it trains the final models using this set and the union of the train and validation splits, evaluating the final models against the test set. 

\subsubsection{Evaluation metrics} Recommenders are evaluated using the original accuracy metrics (PREC, REC, \gls{mrr}, and \gls{ndcg}) and against the following \emph{beyond-accuracy} metrics: novelty  \cite{beyond-accuracy/diversity-mean-inter-list}, item coverage (Cov. Item, quota of recommended items), and distributional diversity (Div. MIL \cite{beyond-accuracy/diversity-mean-inter-list} and Div. Gini \cite{DBLP:journals/tkde/AdomaviciusK12/beyond-accuracy/diversity-gini}). Using these new metrics provides a broader picture of the quality of all recommenders.

\subsubsection{Baselines} Due to space limitations, this work provides only a list of baseline recommenders. A thorough description of all baselines, and the list, range and distribution of their hyper-parameters are in \cite{a-troubling-analysis-of-reproducbility-and-progress-in-recsys-research}. The baselines list is the following: \textbf{Top Popular} \cite{a-troubling-analysis-of-reproducbility-and-progress-in-recsys-research} as a non-personalized approach.
\textbf{UserKNN CF} and \textbf{ItemKNN CF} \cite{a-troubling-analysis-of-reproducbility-and-progress-in-recsys-research} as neighborhood-based CF (similarities: cosine, dice, jaccard, asymmetric cosine, and tversky) and shrinkage term. 
\textbf{RP3beta} \cite{rp3-beta-blockbusters-and-wallflowers-accurate-diverse-and-scalable-recommendations-with-random-walks} as a graph-based approach. 
\textbf{PureSVD} \cite{performace-of-recommender-algorithms-on-top-n-recommendation-tasks} and \textbf{MF BPR} \cite{bpr-bayesian-personalized-ranking-from-implicit-feedback} as matrix factorization models. 
\textbf{SLIM ElasticNet} \cite{a-troubling-analysis-of-reproducbility-and-progress-in-recsys-research,slim} as a machine learning approach.
Lastly, \textbf{EASE R} as a fast linear autoencoder \cite{ease-r-embarrassingly-shallow-autoencoders-for-sparse-data}.

\begin{table}[t]
    \centering
    \caption{Hyper-parameters for \gls{cfgan}. These are divided in two groups. The first group contains specific hyper-parameters of \gls{cfgan}. The second group are hyper-parameters of the generator and discriminator neural networks, values between networks can be different.}
    \label{tab:hyper-parameters}
    \begin{minipage}[c]{\linewidth}
        \renewcommand*\footnoterule{}
        \centering
        \begin{tabular}{l|ccc}
            \toprule
            Hyper-Parameter
                & Type
                & Range
                & Distribution \\
            \midrule
            \# of Epochs
                & Integer
                & $200-400$\footnote{Due to how the training is performed, this range is close to the $1.000$ and $1.500$ epochs used in the reference article}
                & early-stopping \\
            \gls{zr} Coefficient
                & Real
                & $0-1$
                & uniform \\
            \gls{zr} Ratio
                & Integer
                & $10-90$
                & uniform \\
            \gls{pm} Ratio
                & Integer
                & $10-90$
                & uniform \\
            \midrule
            \# of Hidden Layers
                & Integer
                & $1-4$
                & uniform \\
            \# of Hidden Features
                & Integer
                & $50-300$
                & uniform \\
            \# of Steps
                & Integer
                & $1-4$
                & uniform \\
            $l2$ Regularization
                & Real
                & $1\cdot10^{-4} - 1\cdot10^{-1}$
                & log-uniform \\
            Learning Rate
                & Real
                & $1\cdot10^{-4} - 5\cdot10^{-3}$
                & log-uniform \\
            Batch Size
                & Integer
                & $32-256$
                & uniform \\
            Optimizer
                & Categorical
                & ADAM~\cite{adam-a-method-for-stochastic-optimization}
                & - \\
            Activation
                & Categorical
                & sigmoid
                & - \\
            \bottomrule
        \end{tabular}
    \end{minipage}
    \renewcommand*\footnoterule{\bk}
\end{table}

\subsubsection{CFGAN recommenders} The hyper-parameter tuning is done on a total of 18 different \gls{cfgan} models: three datasets (Ciao, ML100K, and ML1M), two modes (item-based \textit{i} and user-based \textit{u}), and three variants (\gls{zr}, \gls{pm}, and \gls{zp}).

To ensure a clear stopping criteria and a fair training for \gls{cfgan}, it is trained using the early-stopping criteria defined in \cite{a-troubling-analysis-of-reproducbility-and-progress-in-recsys-research} and presented in \autoref{sec:theoretical-and-methodological-issues}. The number of minimum and maximum epochs is in \autoref{tab:hyper-parameters}. The early-stopping selects the best number of epochs by using the validation data. The optimal number of epochs is used to train the final model. We recall that the original description of \gls{cfgan} \emph{does not provide} an early-stopping mechanism.

\autoref{tab:hyper-parameters} lists all hyper-parameters of \gls{cfgan}, where hyper-parameters like \emph{optimizer}, \emph{activation} are left unchanged with respect to the reference article. Apart from the number of training epochs, the optimizer, and activation, the rest of the hyper-parameters are set by the Bayesian Search.
\section{Experiments Results \& Discussion}
\label{sec:results-and-discussion}

\begin{table}[t]
    \centering
    \caption{Comparison between the accuracy metrics in the reference article~\cite{cfgan} and those obtained in the replicability experiment (see \autoref{subsec:results-and-discussion:replicability-and-stability}) at recommendation list length of 20. Statistics calculated over 30 executions, evaluating on the last epoch using recommendation lists of length 20. We consistently obtain \emph{lower} results across the three datasets on average. For the Ciao dataset, the original source code trains a different variant (in bold) than the reported in the reference article.
    }
    \label{tab:replicability}
    \begin{minipage}[c]{\linewidth}
        \centering
        \begin{tabular}{lll|cc}
            \toprule
            Dataset & Variant & Stats & PREC & NDCG \\
            \midrule
            Ciao
                & \textbf{iZR}
                & $\text{Mean} \pm \text{Std}$
                & $0.0402 \pm 0.0014$
                & $0.1135 \pm 0.0038$ \\
                & iZP
                & Reference~\cite{cfgan}
                & 0.0450
                & 0.1240 \\
            \midrule
            ML100K
                & iZP
                & $\text{Mean} \pm \text{Std}$
                & $0.2851 \pm 0.0025$
                & $0.4207 \pm 0.0048$ \\
                & iZP
                & Reference~\cite{cfgan}
                & 0.2940
                & 0.4330 \\
            \midrule
            ML1M
                & iZP
                & $\text{Mean} \pm \text{Std}$
                & $0.3079 \pm 0.0011$
                & $0.4035 \pm 0.0016$ \\
                & iZP
                & Reference~\cite{cfgan}
                & 0.3090
                & 0.4060 \\
            \bottomrule
        \end{tabular}
    \end{minipage}
\end{table}

\subsection{RQ1: CFGAN Replicability \& Numerical Stability}
\label{subsec:results-and-discussion:replicability-and-stability}

To address \rqone, we report the results of the replication study, as described in \autoref{subsec:experimental-methodology:original-code-execution}, by using the original source code and data.
This experiment has two goals: \begin{enumerate*}[label=(\roman*)]
    \item verify that published results are replicable; and
    \item measure the numerical stability of \gls{cfgan} given the stochastic nature of its architecture \cite{seed-2-effects-of-random-seeds-on-the-accuracy-of-convolutional-neural-networks,seed-1-on-model-stability-as-a-function-of-random-seed}
\end{enumerate*}.

\autoref{tab:replicability} shows the results of the experiment, we only report two metrics due to space limitations
\footnote{A table with all metrics is available in the supplemental materials of this work.}.
The results reported in the reference article are denoted as \emph{Reference}. Due to the stochastic nature of \gls{cfgan} models, we do not expect to achieve \emph{exact} numerical replicability. 
For all datasets, we see that the replicated results are \textit{lower} than those reported in the reference article. 
For the ML1M dataset, the difference between the average and reported NDCG is $-0.62\%$.
On the smaller ML100K, the results are more varied: $-2.84\%$ between the average and reported NDCG.
For the Ciao dataset, the results could not be replicated due to two factors:  \begin{enumerate*}[label=(\roman*)]
    \item the original source code trained a different variant (iZR) than the reported in the reference article (iZP); and
    \item lack of reproducible hyper-parameters sets for this dataset in the reference article
\end{enumerate*}.
Lastly, with respect to the numerical stability, under \replications executions of this replication, the results indicate that the reference implementation of \gls{cfgan} is numerically stable.

\subsection{RQ2: Impact of Theoretical and Methodological Concerns}
This section reports the results of the experiments related to \rqtwo, those used to measure the impact of the theoretical and methodological concerns raised in \autoref{sec:theoretical-and-methodological-issues}. \autoref{tab:concerns} compares the results of the reference \gls{cfgan} (denoted as Reference), the models tuned in \autoref{subsec:results:reproducibility} (presented in \autoref{tab:reproducibility}), and the variants of this experiment.

\subsubsection{Impact of random noise} As seen in \autoref{sec:generative-adversarial-networks}, \gls{cfgan} receives random noise as part of its input. 
However, in the experiments of the reference article, the random noise is removed. 
This experiment included three different sizes of random noise. 
The results indicate that the recommendation quality improves slightly by removing the random noise, however, as stated in \autoref{sec:theoretical-and-methodological-issues}, it comes at the cost of risking lower generalizability and lower robustness of the generator in a practical use case. 
We argue the random noise should always be present. However, we recall that doing an exhaustive analysis of the impact of random noise in \gls{gan} and \gls{cfgan} is beyond the scope of this paper.

\subsubsection{Impact of condition vector} Similarly as before, in the experiments of the reference article, the condition vector is set to be the user/item profiles, which increases the risk of reaching a trivial solution.
This experiment changed the condition vector to be the user/item classes.
The results show that changing the condition vector with the current \gls{cfgan} architecture dramatically lowers the model's ability to learn to generate accurate profiles. This constitutes a \emph{negative result}, as that the current architecture does not appear to be suitable to handle the user/item classes as the condition vector. 
Identifying an appropriate architecture to do so and an appropriate condition vector to use in scenarios where only past user interactions are available is an open research question that goes beyond the scope of this paper.

\subsubsection{Impact of early-stopping} The reference article does not provide an early-stopping mechanism for \gls{cfgan}, although models in Recommender Systems typically benefit from one, as discussed in \autoref{sec:theoretical-and-methodological-issues}.
This experiment removed the early-stopping and set the maximum number of epochs as 400 (this is the maximum number of epochs set for the early-stopping as seen in \autoref{tab:hyper-parameters}).
Results show that using early-stopping slightly decreases the recommendation quality of \gls{cfgan}, however, we argue that the benefits of using it outweigh the downsides of it, especially if scalability is taken into account. For instance, the iZP variant trains on 645 and 1200 epochs with and without early-stopping, respectively, \idest a decrease of $46.25\%$ in training time and $4.47\%$ in NDCG.

\begin{table}[t]
    \centering
    \caption{Accuracy and beyond-accuracy values for different \gls{cfgan} models for the ML1M dataset at recommendation list length of 20.
    The suffix Reference is the model in the reference article (where $-$ denotes non published values).
    The suffix ES indicates that the model uses early-stopping (see \autoref{tab:reproducibility}), NO-ES indicates it does not.
    The suffix CC indicates that the model uses the user/item class as the condition vector.
    The suffix RN-X means that the model uses random noise of size X.
    Hyper-parameter sets of variants are chosen as described in \autoref{subsec:experimental-methodology:hyper-parameter-tuning} except for those with the Reference suffix.
    }
    \label{tab:concerns}
    \begin{tabular}{l|ccc|l|ccc}
        \toprule
        Variant
            & PREC
            & NDCG
            & \begin{tabular}{@{}c@{}} Cov. \\ Item \end{tabular}
            & Variant
            & PREC
            & NDCG
            & \begin{tabular}{@{}c@{}} Cov. \\ Item \end{tabular} \\
        \midrule
        iZP Reference~\cite{cfgan}
            & 0.3090
            & 0.4060
            & $-$
            & uZP Reference~\cite{cfgan}
            & $-$
            & $-$
            & $-$ \\
        \midrule
        iZP ES
            & 0.2407
            & 0.2972
            & 0.4894
            & uZP ES
            & 0.2764
            & 0.3620
            & 0.1833 \\
        \midrule
        iZP NO-ES
            & 0.2494
            & 0.3111
            & 0.4041
            & uZP NO-ES
            & 0.2797
            & 0.3639
            & 0.1882 \\
        iZP CC
            & 0.0384
            & 0.0507
            & 0.0296
            & uZP CC
            & 0.0916
            & 0.1106
            & 0.0231 \\
        iZP RN-3020
            & 0.2059
            & 0.2475
            & 0.3995
            & uZP RN-1841
            & 0.2737
            & 0.3591
            & 0.1841 \\
        iZP RN-6040
            & 0.1683
            & 0.2000
            & 0.4663
            & uZP RN-3682
            & 0.2781
            & 0.3651
            & 0.1839 \\
        iZP RN-12080
            & 0.1304
            & 0.1471
            & 0.5076
            & uZP RN-7364
            & 0.2759
            & 0.3626
            & 0.1955 \\
        \bottomrule
    \end{tabular}
\end{table}

\begin{table*}[t]
    \centering
    \caption{Accuracy and beyond-accuracy metrics for tuned baselines and \gls{cfgan} on the ML1M dataset at recommendation list length of 20. Higher accuracy values than \gls{cfgan} models reached by baselines in bold. ItemKNN and UserKNN use asymmetric cosine. \gls{cfgan} results are different than \autoref{tab:replicability} due to the hyper-parameter tuning. CFGAN models use early-stopping.}
    \label{tab:reproducibility}
    \begin{tabular}{l|cccc|cccc}
        \toprule
        {}
            & PREC
            & REC
            & MRR
            & NDCG
            & Novelty
            & \begin{tabular}{@{}c@{}} Cov. \\ Item \end{tabular}
            & \begin{tabular}{@{}c@{}} Div. \\ MIL  \end{tabular}
            & \begin{tabular}{@{}c@{}} Div. \\ Gini \end{tabular} \\
        \midrule
        Random          & 0.0099          & 0.0056          & 0.0326          & 0.0108          & 0.0732 & 1.0000 & 0.9946 & 0.8977 \\
        TopPop          & 0.1552          & 0.1146          & 0.3852          & 0.1938          & 0.0473 & 0.0299 & 0.4529 & 0.0095 \\
        \midrule
        UserKNN CF      & 0.2891          & \textbf{0.2570} & \textbf{0.6595} & \textbf{0.3888} & 0.0513 & 0.3286 & 0.8921 & 0.0655 \\
        ItemKNN CF      & 0.2600          & 0.2196          & 0.6254          & 0.3490          & 0.0497 & 0.2097 & 0.8148 & 0.0362 \\
        RP3beta         & 0.2758          & 0.2385          & \textbf{0.6425} & 0.3700          & 0.0506 & 0.3427 & 0.8565 & 0.0528 \\
        PureSVD         & 0.2913          & 0.2421          & \textbf{0.6333} & 0.3783          & 0.0516 & 0.2439 & 0.9142 & 0.0712 \\
        SLIM ElasticNet & \textbf{0.3119} & \textbf{0.2695} & \textbf{0.6724} & \textbf{0.4123} & 0.0514 & 0.3153 & 0.8984 & 0.0696 \\
        MF BPR          & 0.2485          & 0.2103          & 0.5753          & 0.3242          & 0.0512 & 0.3126 & 0.8855 & 0.0631 \\
        EASE R          & \textbf{0.3171} & \textbf{0.2763} & \textbf{0.6795} & \textbf{0.4192} & 0.0518 & 0.3338 & 0.9146 & 0.0803 \\
        \midrule
        CFGAN iZR             & 0.2862          & 0.2547          & 0.6312          & 0.3770          & 0.0542 & 0.4123 & 0.9583 & 0.1459 \\
        CFGAN iPM             & 0.2505          & 0.1950          & 0.5454          & 0.3138          & 0.0523 & 0.3669 & 0.9218 & 0.0901 \\
        CFGAN iZP             & 0.2407          & 0.1742          & 0.5230          & 0.2972          & 0.0530 & 0.4894 & 0.9256 & 0.0901 \\
        \midrule
        CFGAN uZR             & 0.2955          & 0.2473          & 0.6222          & 0.3799          & 0.0523 & 0.2167 & 0.9205 & 0.0837 \\
        CFGAN uPM             & 0.2367          & 0.1928          & 0.5513          & 0.3054          & 0.0516 & 0.1782 & 0.8962 & 0.0550 \\
        CFGAN uZP             & 0.2764          & 0.2342          & 0.6208          & 0.3620          & 0.0513 & 0.1833 & 0.9062 & 0.0617 \\
        \bottomrule
    \end{tabular}
\end{table*}

\subsection{RQ3: Reproducibility Evaluation Against Properly Tuned Baselines}
\label{subsec:results:reproducibility}
To address \rqthree, we report the recommendation quality of \gls{cfgan} and baseline recommenders using a Bayesian hyper-parameter tuning approach, as described in \autoref{subsec:experimental-methodology:hyper-parameter-tuning}. The goal is to evaluate \cite{cfgan} on the same top-N recommendation scenario of the reference paper against a set of properly tuned baselines on accuracy and beyond-accuracy metrics and study if published results are reproducible. 

\autoref{tab:reproducibility} shows the results of accuracy and beyond-accuracy metrics of properly tuned recommenders. Due to space constraints, the focus of this discussion is on the dataset with the highest number of interactions studied in the reference article \cite{cfgan}, \idest ML1M. Results with other datasets are comparable~\footnote{The full results are in the supplemental material~\cite{supplemental_material}.}.

The results indicate that \gls{cfgan} is outperformed by three simple baselines in NDCG, sometimes by almost $10\%$, in particular by other autoencoder-based recommendation models like EASE R and SLIM Elastic Net. These findings are consistent to those reported in several other evaluation studies \cite{are-we-really-making-much-progress-a-worrying-analysis-of-recent-neural-recommendation-approaches,a-troubling-analysis-of-reproducbility-and-progress-in-recsys-research,critically-examining-the-claimed-value-of-convolutions-over-user-item-embedding-maps-for-recommender-systems,evaluation-of-session-bases-recommendation-algorithms,improvements-that-dont-add-up-ad-hoc-retrieval-results-since-1998}. The accuracy across \gls{cfgan} models varies depending on the \gls{cfgan} mode and variant. For instance, the most and least accurate variants are uZR and iZP, respectively, with approximately $21.76\%$ difference in their NDCG metrics. Under the current methodology, we cannot confirm the claim that item-based models or ZP variants outperform other variants, as indicated in the reference article~\cite{cfgan}. In fact, our most accurate variant is uZR.
When looking at beyond-accuracy metrics, item-based \gls{cfgan} models have equal or higher diversity than baselines. In particular, iZR has the highest novelty, item coverage, and distributional diversity, while also being the second-most accurate variant with respect to NDCG. User-based \gls{cfgan} models have less coverage than all baselines. 

It can be seen that the results of the replicability study using hyper-parameter optimization and early-stopping reported in \autoref{tab:reproducibility} are lower than those reported in the replication study in \autoref{tab:replicability}. This indicates that the non-reproducible hyper-parameter search and early-stopping criteria have an important impact on the recommendation quality. 
As a last observation, using the results reported in the reference article \gls{cfgan} would not be competitive against the baselines.

\subsubsection{Scalability}
\label{subsec:experiments-results-and-discussion:scalability}

Concerning the recommendation time, all algorithms are able to create recommendations lists to all users in a total time between 7 and 20 seconds. Differently from other neural models \cite{a-troubling-analysis-of-reproducbility-and-progress-in-recsys-research}, \gls{cfgan} models provide fast recommendations. Due to the lack of random noise, they generate static recommendation lists. 

Concerning the training time, \gls{cfgan} models take more time to train than any baseline. We categorize models into three groups: \begin{enumerate*}[label=(\roman*)]
    \item ItemKNN, UserKNN, PureSVD, RP3beta, and EASE R take between 2 and 25 seconds on average;
    \item machine learning approaches, \idest SLIM and MF BPR take between 3 and 9 minutes to train on average; and 
    \item all \gls{cfgan} models take between 25 and 40 minutes to train on average
\end{enumerate*}. Even on a comparatively small dataset as ML1M, the difference in training time between the first and the last group is two orders of magnitude. Using more performing hardware, \idest GPU could reduce this gap. 

Under this offline evaluation, which is the same as in the original article~\cite{cfgan}, \gls{cfgan} does not generate more accurate recommendations than simple baselines. As \gls{cfgan} is a neural approach, bigger datasets with more complex relations between users, items, and their interactions might increase the accuracy of \gls{cfgan}. However, this is unpractical due to the higher computational cost of \gls{cfgan} models, therefore, we do not report experiments with bigger datasets. 

\section{Conclusions}
\label{sec:conclusion}

This work presents an evaluation study of the family of models of \gls{cfgan}, addressing three research questions under the same top-N recommendation scenario as the reference article \cite{cfgan}. Are previously published results of \gls{cfgan} replicable? What is the impact of the differences between the CFGAN framework and the model evaluated in the reference article? Are previously published results of \gls{cfgan} reproducible? 
Regarding the model's architecture, using as condition vector the user profile and removing the random noise leaves the model prone to a trivial and not useful solution in which the generator behaves as a simple autoencoder, negatively affecting the model's ability to generalize. Due to this, we argue a different approach should be used, which is still an open research question.
The experimental results indicate that \gls{cfgan} is replicable and numerically stable, but not reproducible as it can be outperformed by simple but properly tuned baselines.
This result adds to the recent evidence that properly tuned baselines can outperform complex methods and suggest \gls{cfgan} is not yet a mature recommendation algorithm.

\appendix
\section{Generative Adversarial Networks}
\label{appendix:sec:generative-adversarial-networks}
\glspl{gan} have been successfully applied to numerous prediction and classification tasks. In this work, we discuss a family of generative models originated from \gls{cfgan} used in Recommender Systems. Briefly, a \gls{gan} consists of an adversarial setting between two neural networks that are trained together until they reach convergence. The first neural network is called the \emph{generator}, and it is denoted as \gen. The second network is called the \emph{discriminator} and it is denoted as \disc \cite{gan-communications-acm,gan,gans-an-overview}. 

We use an example to explain the goals of a \gls{gan}. Let us suppose that \gen is a counterfeit organization trying to produce fake bills and that \disc is the local police department in charge of distinguishing fake from real bills. Let us also suppose that after the classification of bills, both the police department and the counterfeiters can know if the classification was accurate or not. The first goal is that \gen learns to generate fake bills as realistically as possible to deceive \disc. The second goal is for \disc to keep up to date with the counterfeited bills not to enter the economy. On every iteration of this setup, \gen updates its counterfeiting methods by looking at the number of errors done by the police. \disc, on the other hand, updates its detection methods, so the identical bills are not misclassified again. This adversarial setting stops when \gen produces such realistic bills that the discriminator can not classify the source of the bills anymore.

\subsection{Theoretical Formulation}

Formally, the data is drawn from a distribution \pdata, and \z is a vector drawn from a prior distribution \pz. The generator \gen is defined as a differentiable function \genfunc with parameters \genparam such that $\x = \genfunc$ meaning that it is a function that maps samples \z drawn from \pz to values \x drawn from a distribution \pg. The learning objective for \gen is to learn a mapping such that $\pg = \pdata$, \idest \gen learns to generate samples drawn from the same distribution as those of the real data.

On the other hand, the discriminator \disc is a function with parameters \discparam such that $y = \disc(\x)$ is a scalar that represents the probability of \x to be drawn from \pdata instead of \pg. The learning objective for \disc is to learn a mapping that assigns high probabilities to samples from \pdata and low probabilities to samples from \pg.

\begin{figure}[t]
    \centering
    \includegraphics[width=0.9\linewidth]{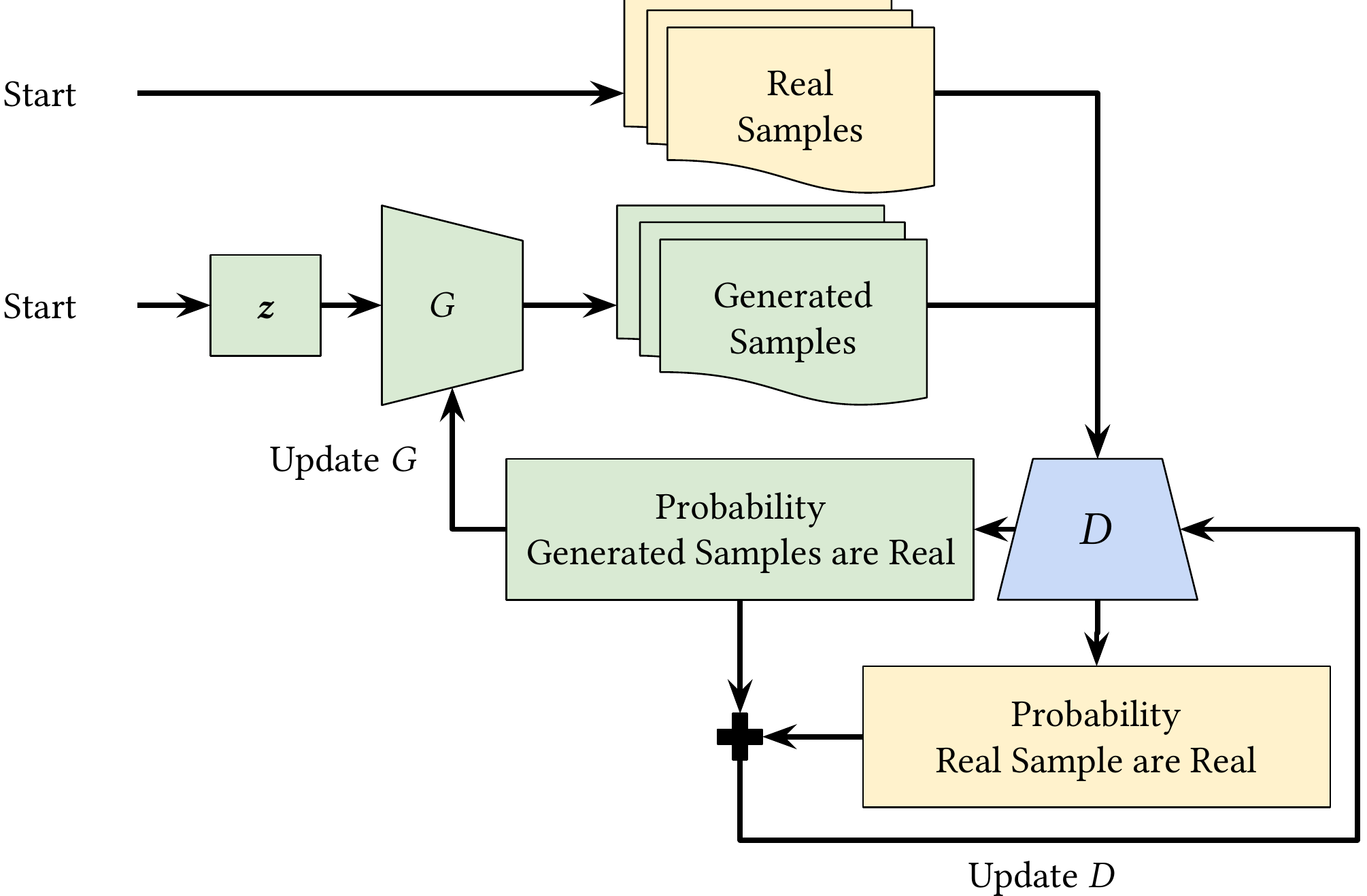}
    \caption{Training process of a \gls{gan}. \gen, \disc, \z are the generator network, discriminator network, and random noise, respectively.}
    \label{fig:gan-training}
\end{figure}

Both \gen and \disc are set up in an adversarial setting where the former tries to maximize the probability of \disc to label generated data as real, thus $ \max_{\genparam} \disc(\genfunc)$ where $\z \sim \pz$ and $\genfunc \sim \pg$. The latter, instead, tries to maximize the probability to distinguish real from generated data, thus $\max_{\discparam} \discfunc$ where $\x \sim \pdata$. In \autoref{fig:gan-training} we illustrate this adversarial setting. Also in \autoref{eq:gan-objective-function} we show the objective function of a \gls{gan} \cite{gan}.

\begin{equation}
    \label{eq:gan-objective-function}
    \max_{\genparam} \max_{\discparam} \ V(\gen, \disc) = \mathbb{E}_{\z \sim \pz} \log \disc(\genfunc)) + \mathbb{E}_{\x \sim \pdata} \log \disc(\x)
\end{equation}

In practice, \glspl{gan} are usually trained to minimize a loss function \ganloss, using \gls{sgd} while translating the expected values $\mathbb{E}$ to cross-entropy losses \cite{cfgan,gan,gan-communications-acm}.

\subsection{Conditional GANs}

The main drawback of a \gls{gan} is that there is no control over the generated samples, \eg in \cite{gan} a \gls{gan} was trained to generate digits from the MNIST dataset. however, it did not control \emph{which} digits were generated by it. \gls{cgan} is an extension of the \gls{gan} model that solves this issue by including a \emph{condition vector} to the generator and discriminator. This vector represents the features or attributes that the generated and discriminated sample must have \cite{conditional-gan}. For example, a work on the MNIST dataset showed that by providing the digit class, \idest ``0'', ``1'', up until ``9'', as a condition to the generator and discriminator a \gls{cgan} generates samples of those digits \cite{conditional-gan}. On \cite{image-to-image-conditional-gan}, a black and white image was used as the condition vector so the generator could generate a colorized version of that image. 

The training procedure of \glspl{cgan} is similar to that of \glspl{gan}. The only difference is that both the generator and discriminator receive a new vector \cond, the condition vector, as part of their input. 
\begin{landscape}

\section{Results RQ1: CFGAN Replicability \& Numerical Stability}

\begin{table}[H]
    \centering
    \caption{Comparison between the accuracy metrics in the reference article~\cite{cfgan} and those obtained in the replicability experiment (see \autoref{subsec:results-and-discussion:replicability-and-stability}) at recommendation list length of 5 and 20. Statistics calculated over 30 executions, evaluating on the last epoch using recommendation lists of length 20. We consistently obtain \emph{lower} results across the three datasets on average. For the Ciao dataset, the original source code trains a different variant (in bold) than the reported in the reference article.}
    \label{appendix:tab:cfgan-replicability}
    \begin{tabular}{llll|cccc}
    \toprule
    @
    & Dataset
    & Variant
    & Stats
    & PREC
    & REC
    & MRR
    & NDCG \\
    \midrule
    
    \multirow{6}{*}{5} 
    & \multirow{2}{*}{Ciao} 
    & \textbf{iZR} & $\text{Mean} \pm \text{Std}$
    & $0.0607 \pm 0.0026$
    & $0.0693 \pm 0.0053$
    & $0.1345 \pm 0.0059$
    & $0.0795 \pm 0.0039$ \\
    
    & & iZP & Reported \cite{cfgan}
    & 0.0720
    & 0.0810
    & 0.1540
    & 0.0920 \\
    \cmidrule{2-8}
        
    & \multirow{2}{*}{ML100K}  
    & iZP & $\text{Mean} \pm \text{Std}$
    & $0.4244 \pm 0.0064$
    & $0.1434 \pm 0.0034$
    & $0.6602 \pm 0.0081$
    & $0.4588 \pm 0.0062$ \\
    
    & & iZP & Reported \cite{cfgan}
    & 0.4440
    & 0.1520
    & 0.6810
    & 0.4760 \\
    \cmidrule{2-8}
    
    & \multirow{2}{*}{ML1M} 
    & iZP & $\text{Mean} \pm \text{Std}$
    & $0.4287 \pm 0.0013$
    & $0.1060 \pm 0.0008$
    & $0.6424 \pm 0.0025$
    & $0.4521 \pm 0.0013$  \\
    
    & & iZP & Reported \cite{cfgan}
    & 0.4320
    & 0.1080
    & 0.6470
    & 0.4550 \\
    \midrule
    
    \multirow{6}{*}{20} 
    & \multirow{2}{*}{Ciao} 
    & \textbf{iZR}
    & $\text{Mean} \pm \text{Std}$
    & $0.0402 \pm 0.0014$ & $0.1788 \pm 0.0071$ & $0.1594 \pm 0.0055$ & $0.1135 \pm 0.0038$ \\
    
    & & iZP & Reported \cite{cfgan}
    & 0.0450              & 0.1940              & 0.1670              & 0.1240              \\
    \cmidrule{2-8}
    
    & \multirow{2}{*}{ML100K} & iZP & $\text{Mean} \pm \text{Std}$
    & $0.2851 \pm 0.0025$ & $0.3400 \pm 0.0050$ & $0.6732 \pm 0.0077$ & $0.4207 \pm 0.0048$ \\
    
    & & iZP & Reported \cite{cfgan}
    & 0.2940              & 0.3600              & 0.6930              & 0.4330              \\
    \cmidrule{2-8}
    
    & \multirow{2}{*}{ML1M}   & iZP & $\text{Mean} \pm \text{Std}$
    & $0.3079 \pm 0.0011$ & $0.2671 \pm 0.0020$ & $0.6566 \pm 0.0024$ & $0.4035 \pm 0.0016$ \\
    
    & & iZP & Reported \cite{cfgan}
    & 0.3090              & 0.2720              & 0.6600              & 0.4060              \\
    \bottomrule
\end{tabular}
\end{table}

\clearpage

\section{Results RQ2: Impact of Theoretical and Methodological Concerns}

\begin{longtable}{l|cccccc|ccccc}
    \caption[]{Accuracy and beyond-accuracy values for different \gls{cfgan} models for the Ciao dataset at recommendation list length of 20.
    The suffix Reference is the model in the reference article (where $-$ denotes non published values).
    The suffix ES indicates that the model uses early-stopping, NO-ES indicates it does not.
    The suffix CC indicates that the model uses the user/item class as the condition vector.
    The suffix RN-X means that the model uses random noise of size X.\label{appendix:tab:ciao-concerns-at-20}} \\
    \toprule
    {}
    & PREC
    & REC
    & MAP
    & MRR
    & NDCG
    & F1
    & Novelty
    & \begin{tabular}{@{}c@{}}
          Div. \\ MIL
    \end{tabular}
    & \begin{tabular}{@{}c@{}}
          Cov. \\ Item
    \end{tabular}
    & \begin{tabular}{@{}c@{}}
          Div. \\ Gini
    \end{tabular}
    & \begin{tabular}{@{}c@{}}
          Div. \\ Shannon
    \end{tabular}         \\
    \midrule
    \endfirsthead
    
    \caption{cont...}\\
    \toprule
    {}
    & PREC
    & REC
    & MAP
    & MRR
    & NDCG
    & F1
    & Novelty
    & \begin{tabular}{@{}c@{}}
          Div. \\ MIL
    \end{tabular}
    & \begin{tabular}{@{}c@{}}
          Cov. \\ Item
    \end{tabular}
    & \begin{tabular}{@{}c@{}}
          Div. \\ Gini
    \end{tabular}
    & \begin{tabular}{@{}c@{}}
          Div. \\ Shannon
    \end{tabular}         \\
    \midrule
    \endhead
    
    iZR Reference~\cite{cfgan}
            & $-$
            & $-$
            & $-$
            & $-$
            & $-$
            & $-$
            & $-$
            & $-$
            & $-$
            & $-$
            & $-$ \\
    \midrule
    iZR ES            & 0.0414 & 0.1783 & 0.0547 & 0.1586 & 0.1149 & 0.0672 &  0.1409 &                                           0.9510 &                                            0.6726 &                                            0.2955 &                                               9.0534 \\
    \midrule
    iZR NO-ES         & 0.0359 & 0.1647 & 0.0498 & 0.1573 & 0.1058 & 0.0589 & 0.1437 & 0.9628 & 0.7869 & 0.3842 & 9.4161 \\
    iZR CC            & 0.0064 & 0.0246 & 0.0045 & 0.0219 & 0.0143 & 0.0102 & 0.1473 & 0.0662 & 0.0423 & 0.0164 & 4.5218 \\
    iZR RN-331        & 0.0393 & 0.1761 & 0.0536 & 0.1489 & 0.1116 & 0.0643 & 0.1414 & 0.9519 & 0.7290 & 0.3094 & 9.1185 \\
    iZR RN-662        & 0.0395 & 0.1718 & 0.0465 & 0.1371 & 0.1046 & 0.0643 & 0.1410 & 0.9480 & 0.7491 & 0.2983 & 9.0630 \\
    iZR RN-1324       & 0.0377 & 0.1650 & 0.0495 & 0.1465 & 0.1052 & 0.0613 & 0.1417 & 0.9500 & 0.8018 & 0.3236 & 9.1666 \\
    
    \midrule
    \midrule
    iPM Reference~\cite{cfgan}
            & $-$
            & $-$
            & $-$
            & $-$
            & $-$
            & $-$
            & $-$
            & $-$
            & $-$
            & $-$
            & $-$ \\
    \midrule
    iPM ES            & 0.0353 & 0.1548 & 0.0451 & 0.1317 & 0.0976 & 0.0575 &  0.1391 &                                           0.9012 &                                            0.5679 &                                            0.1844 &                                               8.3364 \\
    \midrule
    iPM NO-ES         & 0.0356 & 0.1662 & 0.0453 & 0.1322 & 0.1005 & 0.0587 & 0.1370 & 0.9120 & 0.6644 & 0.2192 & 8.5638 \\
    iPM CC            & 0.0083 & 0.0309 & 0.0093 & 0.0348 & 0.0214 & 0.0131 & 0.1498 & 0.9602 & 0.9480 & 0.4495 & 9.5812 \\
    iPM RN-331        & 0.0357 & 0.1533 & 0.0458 & 0.1415 & 0.0990 & 0.0579 & 0.1389 & 0.9036 & 0.5865 & 0.1892 & 8.3721 \\
    iPM RN-662        & 0.0368 & 0.1641 & 0.0448 & 0.1329 & 0.1003 & 0.0601 & 0.1387 & 0.9023 & 0.5939 & 0.1926 & 8.3843 \\
    iPM RN-1324       & 0.0346 & 0.1489 & 0.0420 & 0.1339 & 0.0938 & 0.0562 & 0.1394 & 0.9106 & 0.6726 & 0.2189 & 8.5553 \\
    \midrule
    \midrule
    
    iZP Reference~\cite{cfgan}
            & 0.0450
            & 0.1940
            & $-$
            & 0.1670
            & 0.1240
            & $-$
            & $-$
            & $-$
            & $-$
            & $-$
            & $-$ \\
    \midrule
    iZP ES            & 0.0361 & 0.1635 & 0.0503 & 0.1418 & 0.1041 & 0.0591 &  0.1383 &                                           0.9305 &                                            0.5880 &                                            0.2209 &                                               8.6379 \\
    \midrule
    iZP NO-ES         & 0.0336 & 0.1426 & 0.0384 & 0.1246 & 0.0881 & 0.0544 & 0.1401 & 0.9404 & 0.6533 & 0.2566 & 8.8508 \\
    iZP CC            & 0.0055 & 0.0210 & 0.0034 & 0.0175 & 0.0117 & 0.0088 & 0.1482 & 0.9097 & 0.6177 & 0.1932 & 8.4167 \\
    iZP RN-331        & 0.0345 & 0.1576 & 0.0450 & 0.1350 & 0.0980 & 0.0565 & 0.1383 & 0.9303 & 0.6110 & 0.2299 & 8.6842 \\
    iZP RN-662        & 0.0349 & 0.1492 & 0.0426 & 0.1407 & 0.0953 & 0.0566 & 0.1382 & 0.9257 & 0.6095 & 0.2174 & 8.6045 \\
    iZP RN-1324       & 0.0338 & 0.1538 & 0.0484 & 0.1448 & 0.1007 & 0.0554 & 0.1386 & 0.9292 & 0.6236 & 0.2263 & 8.6638 \\
    \midrule
    \midrule
    
    uZR Reference~\cite{cfgan}
            & $-$
            & $-$
            & $-$
            & $-$
            & $-$
            & $-$
            & $-$
            & $-$
            & $-$
            & $-$
            & $-$ \\
    \midrule
    uZR ES      & 0.0401 & 0.1719 & 0.0523 & 0.1625 & 0.1124 & 0.0650 &  0.1312 &                                           0.8208 &                                            0.2079 &                                            0.0801 &                                               7.1658 \\
        \midrule
        uZR NO-ES   & 0.0324 & 0.1386 & 0.0360 & 0.1147 & 0.0840 & 0.0525 & 0.1430 & 0.9402 & 0.5598 & 0.2418 & 8.7633 \\
        uZR CC      & 0.0217 & 0.1019 & 0.0278 & 0.0892 & 0.0624 & 0.0358 & 0.1221 & 0.1153 & 0.0341 & 0.0168 & 4.5713 \\
        uZR RN-673  & 0.0385 & 0.1635 & 0.0499 & 0.1485 & 0.1069 & 0.0623 & 0.1315 & 0.8228 & 0.2138 & 0.0825 & 7.2085 \\
        uZR RN-1347 & 0.0376 & 0.1596 & 0.0457 & 0.1428 & 0.1017 & 0.0608 & 0.1298 & 0.7864 & 0.1782 & 0.0676 & 6.9228 \\
        uZR RN-2694 & 0.0386 & 0.1640 & 0.0460 & 0.1449 & 0.1033 & 0.0625 & 0.1311 & 0.8172 & 0.2027 & 0.0785 & 7.1371 \\
        \midrule
        \midrule
        
        uPM Reference~\cite{cfgan}
            & $-$
            & $-$
            & $-$
            & $-$
            & $-$
            & $-$
            & $-$
            & $-$
            & $-$
            & $-$
            & $-$ \\
        \midrule
        uPM ES      & 0.0273 & 0.1116 & 0.0333 & 0.1049 & 0.0732 & 0.0439 &  0.1306 &                                           0.6338 &                                            0.0913 &                                            0.0377 &                                               6.0393 \\
        \midrule
        uPM NO-ES   & 0.0289 & 0.1279 & 0.0315 & 0.1094 & 0.0767 & 0.0472 & 0.1320 & 0.7352 & 0.1121 & 0.0524 & 6.5022 \\
        uPM CC      & 0.0084 & 0.0240 & 0.0051 & 0.0233 & 0.0153 & 0.0125 & 0.1395 & 0.0565 & 0.0312 & 0.0161 & 4.4677 \\
        uPM RN-673  & 0.0227 & 0.0967 & 0.0287 & 0.0912 & 0.0631 & 0.0367 & 0.1370 & 0.6737 & 0.0891 & 0.0415 & 6.1492 \\
        uPM RN-1347 & 0.0224 & 0.1012 & 0.0259 & 0.0868 & 0.0613 & 0.0367 & 0.1321 & 0.5781 & 0.0616 & 0.0323 & 5.6931 \\
        uPM RN-2694 & 0.0223 & 0.0970 & 0.0282 & 0.0914 & 0.0626 & 0.0362 & 0.1324 & 0.5298 & 0.0683 & 0.0294 & 5.6474 \\
        \midrule
        \midrule
        
        uZP Reference~\cite{cfgan}
            & $-$
            & $-$
            & $-$
            & $-$
            & $-$
            & $-$
            & $-$
            & $-$
            & $-$
            & $-$
            & $-$ \\
        \midrule
        uZP ES      & 0.0230 & 0.1054 & 0.0228 & 0.0664 & 0.0579 & 0.0378 &  0.1204 &                                           0.1238 &                                            0.0341 &                                            0.0169 &                                               4.5934 \\
        \midrule
        uZP NO-ES   & 0.0227 & 0.1088 & 0.0283 & 0.0848 & 0.0646 & 0.0375 & 0.1212 & 0.1209 & 0.0371 & 0.0168 & 4.5927 \\
        uZP CC      & 0.0216 & 0.0989 & 0.0254 & 0.0812 & 0.0593 & 0.0354 & 0.1220 & 0.1159 & 0.0312 & 0.0167 & 4.5733 \\
        uZP RN-673  & 0.0223 & 0.1009 & 0.0278 & 0.0826 & 0.0621 & 0.0365 & 0.1205 & 0.1238 & 0.0349 & 0.0169 & 4.5925 \\
        uZP RN-1347 & 0.0233 & 0.1079 & 0.0310 & 0.0902 & 0.0670 & 0.0383 & 0.1204 & 0.1240 & 0.0364 & 0.0169 & 4.5988 \\
        uZP RN-2694 & 0.0226 & 0.1052 & 0.0265 & 0.0838 & 0.0621 & 0.0372 & 0.1206 & 0.1227 & 0.0371 & 0.0169 & 4.5945 \\
    \bottomrule
\end{longtable}

\clearpage

\begin{longtable}{l|cccccc|ccccc}
    \caption[]{Accuracy and beyond-accuracy values for different \gls{cfgan} models for the ML100K dataset at recommendation list length of 20.
    The suffix Reference is the model in the reference article (where $-$ denotes non published values).
    The suffix ES indicates that the model uses early-stopping, NO-ES indicates it does not.
    The suffix CC indicates that the model uses the user/item class as the condition vector.
    The suffix RN-X means that the model uses random noise of size X.\label{appendix:tab:ml100k-concerns-at-20}} \\
    \toprule
    {}
    & PREC
    & REC
    & MAP
    & MRR
    & NDCG
    & F1
    & Novelty
    & \begin{tabular}{@{}c@{}}
          Div. \\ MIL
    \end{tabular}
    & \begin{tabular}{@{}c@{}}
          Cov. \\ Item
    \end{tabular}
    & \begin{tabular}{@{}c@{}}
          Div. \\ Gini
    \end{tabular}
    & \begin{tabular}{@{}c@{}}
          Div. \\ Shannon
    \end{tabular}         \\
    \midrule
    \endfirsthead
    
    \caption{cont...}\\
    \toprule
    {}
    & PREC
    & REC
    & MAP
    & MRR
    & NDCG
    & F1
    & Novelty
    & \begin{tabular}{@{}c@{}}
          Div. \\ MIL
    \end{tabular}
    & \begin{tabular}{@{}c@{}}
          Cov. \\ Item
    \end{tabular}
    & \begin{tabular}{@{}c@{}}
          Div. \\ Gini
    \end{tabular}
    & \begin{tabular}{@{}c@{}}
          Div. \\ Shannon
    \end{tabular}         \\
    \midrule
    \endhead
    
    iZR Reference~\cite{cfgan}
            & $-$
            & $-$
            & $-$
            & $-$
            & $-$
            & $-$
            & $-$
            & $-$
            & $-$
            & $-$
            & $-$ \\
    \midrule
    iZR ES      & 0.2415 & 0.3025 & 0.1923 & 0.5866 & 0.3546 & 0.2685 &  0.1073 &                                           0.9017 &                                            0.3400 &                                            0.1235 &                                               8.0988 \\
    \midrule
    iZR NO-ES   & 0.2317 & 0.3030 & 0.1821 & 0.5663 & 0.3446 & 0.2626 & 0.1088 & 0.9121 & 0.3956 & 0.1468 & 8.3357 \\
    iZR CC      & 0.0297 & 0.0247 & 0.0100 & 0.0959 & 0.0357 & 0.0270 & 0.1316 & 0.5090 & 0.0417 & 0.0230 & 5.4954 \\
    iZR RN-471  & 0.2401 & 0.3026 & 0.1913 & 0.5815 & 0.3532 & 0.2678 & 0.1068 & 0.8943 & 0.3206 & 0.1132 & 7.9756  \\
    iZR RN-943  & 0.2408 & 0.2998 & 0.1917 & 0.5940 & 0.3546 & 0.2671 & 0.1070 & 0.8992 & 0.3406 & 0.1174 & 8.0295  \\
    iZR RN-1886 & 0.2327 & 0.2923 & 0.1817 & 0.5774 & 0.3414 & 0.2591 & 0.1071 & 0.8979 & 0.3128 & 0.1138 & 7.9833  \\
    \midrule
    \midrule
    
    iPM Reference~\cite{cfgan}
            & $-$
            & $-$
            & $-$
            & $-$
            & $-$
            & $-$
            & $-$
            & $-$
            & $-$
            & $-$
            & $-$ \\
    \midrule
    iPM ES      & 0.2171 & 0.2350 & 0.1641 & 0.5247 & 0.3023 & 0.2257 &  0.1037 &                                           0.8417 &                                            0.2390 &                                            0.0731 &                                               7.3341 \\
    \midrule
    iPM NO-ES   & 0.2466 & 0.2829 & 0.2004 & 0.5749 & 0.3506 & 0.2635 & 0.1056 & 0.8875 & 0.3067 & 0.1025 & 7.8294 \\
    iPM CC      & 0.0128 & 0.0112 & 0.0027 & 0.0294 & 0.0131 & 0.0119 & 0.1498 & 0.5732 & 0.1518 & 0.0283 & 5.7504 \\
    iPM RN-471  & 0.2107 & 0.2213 & 0.1533 & 0.4933 & 0.2853 & 0.2159 & 0.1034 & 0.8364 & 0.1972 & 0.0702 & 7.2708  \\
    iPM RN-943  & 0.2144 & 0.2332 & 0.1618 & 0.5147 & 0.2984 & 0.2234 & 0.1033 & 0.8392 & 0.2257 & 0.0711 & 7.2919  \\
    iPM RN-1886 & 0.2128 & 0.2264 & 0.1573 & 0.5106 & 0.2917 & 0.2194 & 0.1039 & 0.8440 & 0.2129 & 0.0734 & 7.3358  \\
    \midrule
    \midrule
    iZP Reference~\cite{cfgan}
            & 0.2940
            & 0.3600
            & $-$
            & 0.6930
            & 0.4330
            & $-$
            & $-$
            & $-$
            & $-$
            & $-$
            & $-$ \\
    
    \midrule
    iZP ES      & 0.2757 & 0.3283 & 0.2394 & 0.6580 & 0.4041 & 0.2997 &  0.1040 &                                           0.8572 &                                            0.2589 &                                            0.0848 &                                               7.5584 \\
    \midrule
    iZP NO-ES   & 0.2759 & 0.3311 & 0.2386 & 0.6553 & 0.4055 & 0.3010 & 0.1051 & 0.8802 & 0.2934 & 0.1014 & 7.8161 \\
    iZP CC      & 0.1290 & 0.1133 & 0.0638 & 0.2802 & 0.1525 & 0.1206 & 0.1011 & 0.4057 & 0.0799 & 0.0199 & 5.3747 \\
    iZP RN-471  & 0.2750 & 0.3271 & 0.2377 & 0.6589 & 0.4029 & 0.2988 & 0.1038 & 0.8535 & 0.2613 & 0.0835 & 7.5368  \\
    iZP RN-943  & 0.2726 & 0.3212 & 0.2373 & 0.6537 & 0.3999 & 0.2949 & 0.1031 & 0.8329 & 0.2511 & 0.0753 & 7.3841  \\
    iZP RN-1886 & 0.2725 & 0.3220 & 0.2354 & 0.6543 & 0.3992 & 0.2952 & 0.1039 & 0.8533 & 0.2589 & 0.0835 & 7.5360  \\
    \midrule
    \midrule
    
    uZR Reference~\cite{cfgan}
            & $-$
            & $-$
            & $-$
            & $-$
            & $-$
            & $-$
            & $-$
            & $-$
            & $-$
            & $-$
            & $-$ \\
    \midrule
    uZR ES      & 0.2754 & 0.3387 & 0.2409 & 0.6500 & 0.4078 & 0.3038 &  0.1060 &                                           0.8844 &                                            0.2371 &                                            0.1018 &                                               7.8011 \\
    \midrule
    uZR NO-ES   & 0.2732 & 0.3369 & 0.2365 & 0.6507 & 0.4049 & 0.3018 &  0.1085 &                                           0.9158 &                                            0.3085 &                                            0.1388 &                                               8.2464 \\
    uZR CC      & 0.1100 & 0.1520 & 0.0665 & 0.3847 & 0.1689 & 0.1276 &  0.1011 &                                           0.3616 &                                            0.0411 &                                            0.0178 &                                               5.0993 \\
    uZR RN-826  & 0.2768 & 0.3382 & 0.2427 & 0.6533 & 0.4100 & 0.3045 &  0.1065 &                                           0.8922 &                                            0.2414 &                                            0.1091 &                                               7.8922 \\
    uZR RN-1653 & 0.2725 & 0.3360 & 0.2375 & 0.6456 & 0.4036 & 0.3009 &  0.1060 &                                           0.8853 &                                            0.2250 &                                            0.0996 &                                               7.7634 \\
    uZR RN-3306 & 0.2694 & 0.3286 & 0.2336 & 0.6356 & 0.3980 & 0.2961 &  0.1070 &                                           0.8869 &                                            0.2686 &                                            0.1118 &                                               7.9293 \\
    \midrule
    \midrule
    
    uPM Reference~\cite{cfgan}
            & $-$
            & $-$
            & $-$
            & $-$
            & $-$
            & $-$
            & $-$
            & $-$
            & $-$
            & $-$
            & $-$ \\
    \midrule
    uPM ES      & 0.2326 & 0.3057 & 0.1864 & 0.5809 & 0.3498 & 0.2642 &  0.1058 &                                           0.8699 &                                            0.2607 &                                            0.0951 &                                               7.7164 \\
    \midrule
    uPM NO-ES   & 0.2154 & 0.2872 & 0.1586 & 0.5223 & 0.3158 & 0.2462 &  0.1067 &                                           0.8861 &                                            0.3116 &                                            0.1111 &                                               7.9411 \\
    uPM CC      & 0.1529 & 0.1691 & 0.0923 & 0.4148 & 0.2090 & 0.1606 &  0.0978 &                                           0.4355 &                                            0.0478 &                                            0.0201 &                                               5.3440 \\
    uPM RN-826  & 0.2284 & 0.2969 & 0.1823 & 0.5710 & 0.3425 & 0.2582 &  0.1055 &                                           0.8733 &                                            0.2511 &                                            0.0990 &                                               7.7671 \\
    uPM RN-1653 & 0.2426 & 0.3050 & 0.1937 & 0.5717 & 0.3550 & 0.2703 &  0.1054 &                                           0.8711 &                                            0.2529 &                                            0.0976 &                                               7.7464 \\
    uPM RN-3306 & 0.2376 & 0.2929 & 0.1866 & 0.5654 & 0.3448 & 0.2624 &  0.1058 &                                           0.8727 &                                            0.2474 &                                            0.0982 &                                               7.7524 \\
    \midrule
    \midrule
    
    uZP Reference~\cite{cfgan}
            & $-$
            & $-$
            & $-$
            & $-$
            & $-$
            & $-$
            & $-$
            & $-$
            & $-$
            & $-$
            & $-$ \\
    \midrule
    uZP ES      & 0.2555 & 0.3163 & 0.2194 & 0.6399 & 0.3834 & 0.2827 &  0.1055 &                                           0.8855 &                                            0.2529 &                                            0.1055 &                                               7.8565 \\
    \midrule
    uZP NO-ES   & 0.2619 & 0.3289 & 0.2221 & 0.6379 & 0.3909 & 0.2916 &  0.1073 &                                           0.9100 &                                            0.3303 &                                            0.1358 &                                               8.2238 \\
    uZP CC      & 0.1239 & 0.1738 & 0.0798 & 0.3699 & 0.1888 & 0.1446 &  0.0997 &                                           0.3894 &                                            0.0430 &                                            0.0186 &                                               5.1766 \\
    uZP RN-826  & 0.2590 & 0.3225 & 0.2230 & 0.6406 & 0.3892 & 0.2873 &  0.1057 &                                           0.8905 &                                            0.2632 &                                            0.1112 &                                               7.9291 \\
    uZP RN-1653 & 0.2530 & 0.3204 & 0.2149 & 0.6311 & 0.3812 & 0.2828 &  0.1063 &                                           0.8970 &                                            0.2317 &                                            0.1110 &                                               7.9023 \\
    uZP RN-3306 & 0.2449 & 0.3090 & 0.2040 & 0.6191 & 0.3679 & 0.2732 &  0.1058 &                                           0.8812 &                                            0.2214 &                                            0.0997 &                                               7.7605 \\
    \bottomrule
\end{longtable}

\clearpage

\begin{longtable}{l|cccccc|ccccc}
    \caption[]{Accuracy and beyond-accuracy values for different \gls{cfgan} models for the ML1M dataset at recommendation list length of 20.
    The suffix Reference is the model in the reference article (where $-$ denotes non published values).
    The suffix ES indicates that the model uses early-stopping, NO-ES indicates it does not.
    The suffix CC indicates that the model uses the user/item class as the condition vector.
    The suffix RN-X means that the model uses random noise of size X.\label{appendix:tab:ml1m-concerns-at-20}} \\
    \toprule
    {}
    & PREC
    & REC
    & MAP
    & MRR
    & NDCG
    & F1
    & Novelty
    & \begin{tabular}{@{}c@{}}
          Div. \\ MIL
    \end{tabular}
    & \begin{tabular}{@{}c@{}}
          Cov. \\ Item
    \end{tabular}
    & \begin{tabular}{@{}c@{}}
          Div. \\ Gini
    \end{tabular}
    & \begin{tabular}{@{}c@{}}
          Div. \\ Shannon
    \end{tabular}         \\
    \midrule
    \endfirsthead
    
    \caption{cont...}\\
    \toprule
    {}
    & PREC
    & REC
    & MAP
    & MRR
    & NDCG
    & F1
    & Novelty
    & \begin{tabular}{@{}c@{}}
          Div. \\ MIL
    \end{tabular}
    & \begin{tabular}{@{}c@{}}
          Cov. \\ Item
    \end{tabular}
    & \begin{tabular}{@{}c@{}}
          Div. \\ Gini
    \end{tabular}
    & \begin{tabular}{@{}c@{}}
          Div. \\ Shannon
    \end{tabular}         \\
    \midrule
    \endhead
    
    iZR Reference~\cite{cfgan}
            & $-$
            & $-$
            & $-$
            & $-$
            & $-$
            & $-$
            & $-$
            & $-$
            & $-$
            & $-$
            & $-$ \\
    \midrule
    iZR ES      & 0.2862 & 0.2547 & 0.2146 & 0.6312 & 0.3770 & 0.2696 &  0.0542 &                                           0.9583 &                                            0.4123 &                                            0.1459 &                                               9.4737 \\
    \midrule
    iZR NO-ES   & 0.2909 & 0.2608 & 0.2195 & 0.6361 & 0.3833 & 0.2750 &  0.0546 &                                           0.9621 &                                            0.4402 &                                            0.1592 &                                               9.5998 \\
    iZR CC      & 0.0599 & 0.0431 & 0.0237 & 0.1688 & 0.0707 & 0.0501 &  0.0584 &                                           0.8577 &                                            0.7257 &                                            0.1003 &                                               8.4716 \\
    iZR RN-3020  & 0.2861 & 0.2545 & 0.2140 & 0.6267 & 0.3759 & 0.2693 &  0.0542 &                                           0.9588 &                                            0.4058 &                                            0.1454 &                                               9.4711 \\
    iZR RN-6040  & 0.2864 & 0.2538 & 0.2148 & 0.6250 & 0.3758 & 0.2691 &  0.0542 &                                           0.9580 &                                            0.4063 &                                            0.1427 &                                               9.4464 \\
    iZR RN-12080 & 0.2852 & 0.2526 & 0.2140 & 0.6242 & 0.3747 & 0.2679 &  0.0541 &                                           0.9574 &                                            0.3987 &                                            0.1418 &                                               9.4328 \\
    
    \midrule
    \midrule
    iPM Reference~\cite{cfgan}
            & $-$
            & $-$
            & $-$
            & $-$
            & $-$
            & $-$
            & $-$
            & $-$
            & $-$
            & $-$
            & $-$ \\
    \midrule
    iPM ES      & 0.2505 & 0.1950 & 0.1734 & 0.5454 & 0.3138 & 0.2193 &  0.0523 &                                           0.9218 &                                            0.3669 &                                            0.0901 &                                               8.7458 \\
    \midrule
    iPM NO-ES   & 0.2534 & 0.1960 & 0.1763 & 0.5480 & 0.3170 & 0.2211 &  0.0528 &                                           0.9302 &                                            0.3914 &                                            0.0970 &                                               8.8672 \\
    iPM CC      & 0.0145 & 0.0069 & 0.0043 & 0.0601 & 0.0171 & 0.0093 &  0.0696 &                                           0.3734 &                                            0.0179 &                                            0.0084 &                                               5.2277 \\
    iPM RN-3020  & 0.2487 & 0.1888 & 0.1707 & 0.5332 & 0.3085 & 0.2147 &  0.0534 &                                           0.9319 &                                            0.3672 &                                            0.1005 &                                               8.9107 \\
    iPM RN-6040  & 0.2107 & 0.1550 & 0.1398 & 0.4687 & 0.2603 & 0.1786 &  0.0597 &                                           0.9443 &                                            0.4856 &                                            0.1206 &                                               9.1833 \\
    iPM RN-12080 & 0.2285 & 0.1545 & 0.1496 & 0.4767 & 0.2725 & 0.1843 &  0.0594 &                                           0.9609 &                                            0.4294 &                                            0.1390 &                                               9.4311 \\
    \midrule
    \midrule
    
    iZP Reference~\cite{cfgan}
            & 0.3090
            & 0.2720
            & $-$
            & 0.6600
            & 0.4060
            & $-$
            & $-$
            & $-$
            & $-$
            & $-$
            & $-$ \\
    \midrule
    iZP ES      & 0.2407 & 0.1742 & 0.1661 & 0.5230 & 0.2972 & 0.2021 &  0.0530 &                                           0.9256 &                                            0.4894 &                                            0.0901 &                                               8.7580 \\
    \midrule
    iZP NO-ES   & 0.2494 & 0.1888 & 0.1743 & 0.5434 & 0.3111 & 0.2149 &  0.0535 &                                           0.9477 &                                            0.4041 &                                            0.1060 &                                               9.0435 \\
    iZP CC      & 0.0384 & 0.0218 & 0.0166 & 0.1648 & 0.0507 & 0.0278 &  0.0653 &                                           0.5382 &                                            0.0296 &                                            0.0109 &                                               5.6056 \\
    iZP RN-3020  & 0.2059 & 0.1377 & 0.1316 & 0.4490 & 0.2475 & 0.1650 &  0.0535 &                                           0.9459 &                                            0.3995 &                                            0.1034 &                                               9.0063 \\
    iZP RN-6040  & 0.1683 & 0.1112 & 0.0983 & 0.3808 & 0.2000 & 0.1339 &  0.0560 &                                           0.9479 &                                            0.4663 &                                            0.1204 &                                               9.2016 \\
    iZP RN-12080 & 0.1304 & 0.0724 & 0.0670 & 0.2906 & 0.1471 & 0.0931 &  0.0599 &                                           0.9587 &                                            0.5076 &                                            0.1409 &                                               9.4485 \\
    \midrule
    \midrule
    
    uZR Reference~\cite{cfgan}
            & $-$
            & $-$
            & $-$
            & $-$
            & $-$
            & $-$
            & $-$
            & $-$
            & $-$
            & $-$
            & $-$ \\
    \midrule
    uZR ES      & 0.2955 & 0.2473 & 0.2241 & 0.6222 & 0.3799 & 0.2692 &  0.0523 &                                           0.9205 &                                            0.2167 &                                            0.0837 &                                               8.6304 \\
    \midrule
    uZR NO-ES   & 0.2959 & 0.2519 & 0.2281 & 0.6376 & 0.3856 & 0.2721 &  0.0522 &                                           0.9175 &                                            0.2273 &                                            0.0782 &                                               8.5571 \\
    uZR CC      & 0.1459 & 0.1118 & 0.0826 & 0.3695 & 0.1831 & 0.1266 &  0.0475 &                                           0.4457 &                                            0.0291 &                                            0.0093 &                                               5.3905 \\
    uZR RN-1841  & 0.2862 & 0.2414 & 0.2136 & 0.6092 & 0.3683 & 0.2619 &  0.0519 &                                           0.9030 &                                            0.2127 &                                            0.0696 &                                               8.3787 \\
    uZR RN-3682 & 0.2915 & 0.2425 & 0.2222 & 0.6211 & 0.3760 & 0.2648 &  0.0514 &                                           0.8929 &                                            0.2021 &                                            0.0638 &                                               8.2516 \\
    uZR RN-7364 & 0.2902 & 0.2435 & 0.2196 & 0.6154 & 0.3739 & 0.2648 &  0.0511 &                                           0.8797 &                                            0.1988 &                                            0.0572 &                                               8.0952 \\
    \midrule
    \midrule
    
    uPM Reference~\cite{cfgan}
            & $-$
            & $-$
            & $-$
            & $-$
            & $-$
            & $-$
            & $-$
            & $-$
            & $-$
            & $-$
            & $-$ \\
    \midrule
    uPM ES      & 0.2367 & 0.1928 & 0.1629 & 0.5513 & 0.3054 & 0.2125 &  0.0516 &                                           0.8962 &                                            0.1782 &                                            0.0550 &                                               8.0858 \\
    \midrule
    uPM NO-ES   & 0.2402 & 0.2053 & 0.1660 & 0.5594 & 0.3135 & 0.2214 &  0.0512 &                                           0.8767 &                                            0.1877 &                                            0.0507 &                                               7.9488 \\
    uPM CC      & 0.1443 & 0.1025 & 0.0786 & 0.3268 & 0.1720 & 0.1198 &  0.0480 &                                           0.4331 &                                            0.0291 &                                            0.0091 &                                               5.3832 \\
    uPM RN-1841  & 0.2250 & 0.1729 & 0.1495 & 0.5155 & 0.2835 & 0.1956 &  0.0519 &                                           0.8958 &                                            0.1632 &                                            0.0538 &                                               8.0511 \\
    uPM RN-3682 & 0.2297 & 0.1776 & 0.1545 & 0.5250 & 0.2902 & 0.2003 &  0.0513 &                                           0.8677 &                                            0.1749 &                                            0.0473 &                                               7.8506 \\
    uPM RN-7364 & 0.1963 & 0.1529 & 0.1259 & 0.4935 & 0.2529 & 0.1719 &  0.0523 &                                           0.8673 &                                            0.1475 &                                            0.0419 &                                               7.6993 \\
    \midrule
    \midrule
    
    uZP Reference~\cite{cfgan}
            & $-$
            & $-$
            & $-$
            & $-$
            & $-$
            & $-$
            & $-$
            & $-$
            & $-$
            & $-$
            & $-$ \\
    \midrule
    uZP ES      & 0.2764 & 0.2342 & 0.2074 & 0.6208 & 0.3620 & 0.2536 &  0.0513 &                                           0.9062 &                                            0.1833 &                                            0.0617 &                                               8.2408 \\
    \midrule
    uZP NO-ES   & 0.2797 & 0.2323 & 0.2114 & 0.6198 & 0.3639 & 0.2538 &  0.0513 &                                           0.9052 &                                            0.1882 &                                            0.0622 &                                               8.2519 \\
    uZP CC      & 0.0916 & 0.0715 & 0.0399 & 0.2513 & 0.1106 & 0.0803 &  0.0511 &                                           0.3083 &                                            0.0231 &                                            0.0074 &                                               5.0549 \\
    uZP RN-1841  & 0.2737 & 0.2322 & 0.2050 & 0.6171 & 0.3591 & 0.2512 &  0.0513 &                                           0.9050 &                                            0.1841 &                                            0.0611 &                                               8.2269 \\
    uZP RN-3682 & 0.2781 & 0.2357 & 0.2102 & 0.6241 & 0.3651 & 0.2552 &  0.0514 &                                           0.9088 &                                            0.1839 &                                            0.0631 &                                               8.2737 \\
    uZP RN-7364 & 0.2759 & 0.2363 & 0.2075 & 0.6203 & 0.3626 & 0.2546 &  0.0514 &                                           0.9093 &                                            0.1955 &                                            0.0647 &                                               8.3079 \\
    \bottomrule
\end{longtable}

\clearpage

\section{Results RQ3: Reproducibility Evaluation Against Properly Tuned Baselines}
\label{subsec:results-and-experiments:accuracy-and-beyond-accuracy-evaluation}

\begin{longtable}{l|ccccccc|ccccc}
    \caption[]{Accuracy and beyond-accuracy metrics for tuned baselines and \gls{cfgan} on the Ciao dataset at
    recommendation list length of 20. \gls{cfgan} results are different than \autoref{appendix:tab:ciao-concerns-at-20} due to the hyper-parameter tuning.\label{appendix:tab:ciao-reproducibility-@20}} \\
    \toprule
        {}
        & PREC
        & REC
        & MAP
        & MRR
        & NDCG
        & F1
        & ARHR
        & Novelty
        & \begin{tabular}{@{}c@{}} Div. \\ MIL \end{tabular}
        & \begin{tabular}{@{}c@{}}               Cov. \\ Item \end{tabular}
        & \begin{tabular}{@{}c@{}}Div. \\ Gini \end{tabular}
        & \begin{tabular}{@{}c@{}}                        Div. \\ Shannon\end{tabular}       \\
        \midrule
        \endfirsthead
        
    \caption[]{cont...} \\
    \toprule
        {}
        & PREC
        & REC
        & MAP
        & MRR
        & NDCG
        & F1
        & ARHR
        & Novelty
        & \begin{tabular}{@{}c@{}} Div. \\ MIL \end{tabular}
        & \begin{tabular}{@{}c@{}}               Cov. \\ Item \end{tabular}
        & \begin{tabular}{@{}c@{}}Div. \\ Gini \end{tabular}
        & \begin{tabular}{@{}c@{}}                        Div. \\ Shannon\end{tabular}       \\
        \midrule
        \endhead
        
        Random                  & 0.0045 & 0.0164 & 0.0028 & 0.0124 & 0.0092 & 0.0070 & 0.0137 & 0.1603 & 0.9852 & 1.0000 & 0.8229 & 10.3215 \\
        TopPop                  & 0.0232 & 0.1091 & 0.0282 & 0.0829 & 0.0645 & 0.0383 & 0.1021 & 0.1199 & 0.1312 & 0.0356 & 0.0170 & 4.6100 \\
        \midrule
        UserKNN CF cosine       & 0.0416 & 0.1874 & 0.0574 & 0.1650 & 0.1201 & 0.0680 & 0.2087 & 0.1325 & 0.8451 & 0.5078 & 0.1269 & 7.7617 \\
        UserKNN CF dice         & 0.0383 & 0.1747 & 0.0515 & 0.1531 & 0.1102 & 0.0628 & 0.1912 & 0.1317 & 0.8157 & 0.4246 & 0.1030 & 7.4729 \\
        UserKNN CF jaccard      & 0.0388 & 0.1786 & 0.0516 & 0.1527 & 0.1112 & 0.0637 & 0.1914 & 0.1308 & 0.7905 & 0.3764 & 0.0884 & 7.2620 \\
        UserKNN CF asymmetric   & 0.0376 & 0.1689 & 0.0489 & 0.1477 & 0.1062 & 0.0615 & 0.1848 & 0.1307 & 0.7818 & 0.3445 & 0.0836 & 7.1874 \\
        UserKNN CF tversky      & 0.0413 & 0.1850 & 0.0578 & 0.1621 & 0.1193 & 0.0676 & 0.2079 & 0.1323 & 0.8388 & 0.5241 & 0.1274 & 7.7472 \\
        ItemKNN CF cosine       & 0.0410 & 0.1807 & 0.0555 & 0.1589 & 0.1165 & 0.0669 & 0.2022 & 0.1313 & 0.8116 & 0.4566 & 0.1063 & 7.4955 \\
        ItemKNN CF dice         & 0.0387 & 0.1709 & 0.0512 & 0.1495 & 0.1090 & 0.0631 & 0.1898 & 0.1357 & 0.8820 & 0.6585 & 0.1880 & 8.2791 \\
        ItemKNN CF jaccard      & 0.0371 & 0.1611 & 0.0518 & 0.1499 & 0.1069 & 0.0603 & 0.1894 & 0.1356 & 0.8800 & 0.6548 & 0.1858 & 8.2591 \\
        ItemKNN CF asymmetric   & 0.0409 & 0.1792 & 0.0548 & 0.1587 & 0.1157 & 0.0665 & 0.2012 & 0.1312 & 0.8120 & 0.4410 & 0.1050 & 7.4850 \\
        ItemKNN CF tversky      & 0.0375 & 0.1596 & 0.0510 & 0.1511 & 0.1065 & 0.0607 & 0.1899 & 0.1354 & 0.8794 & 0.6578 & 0.1856 & 8.2554 \\
        RP3beta                 & 0.0442 & 0.1971 & 0.0601 & 0.1730 & 0.1262 & 0.0722 & 0.2194 & 0.1355 & 0.8688 & 0.7060 & 0.1868 & 8.2145 \\
        PureSVD                 & 0.0366 & 0.1596 & 0.0443 & 0.1428 & 0.0999 & 0.0595 & 0.1784 & 0.1332 & 0.8561 & 0.3875 & 0.1165 & 7.7012 \\
        SLIM ElasticNet         & 0.0430 & 0.1852 & 0.0582 & 0.1640 & 0.1208 & 0.0698 & 0.2114 & 0.1343 & 0.8758 & 0.5798 & 0.1628 & 8.1157 \\
        MF BPR                  & 0.0246 & 0.1013 & 0.0323 & 0.1052 & 0.0683 & 0.0396 & 0.1271 & 0.1406 & 0.9320 & 0.8641 & 0.3290 & 9.0762 \\
        EASE R                  & 0.0425 & 0.1825 & 0.0560 & 0.1591 & 0.1176 & 0.0689 & 0.2059 & 0.1343 & 0.8799 & 0.5479 & 0.1596 & 8.1095 \\
        \midrule
        CFGAN iZR               & 0.0414 & 0.1783 & 0.0547 & 0.1586 & 0.1149 & 0.0672 & 0.2033 & 0.1409 & 0.9510 & 0.6726 & 0.2955 & 9.0534 \\
        CFGAN iPM               & 0.0353 & 0.1548 & 0.0451 & 0.1317 & 0.0976 & 0.0575 & 0.1659 & 0.1391 & 0.9012 & 0.5679 & 0.1844 & 8.3364 \\
        CFGAN iZP               & 0.0361 & 0.1635 & 0.0503 & 0.1418 & 0.1041 & 0.0591 & 0.1772 & 0.1383 & 0.9305 & 0.5880 & 0.2209 & 8.6379  \\
        \midrule
        CFGAN uZR               & 0.0401 & 0.1719 & 0.0523 & 0.1625 & 0.1124 & 0.0650 & 0.2056 & 0.1312 & 0.8208 & 0.2079 & 0.0801 & 7.1658  \\
        CFGAN uPM               & 0.0273 & 0.1116 & 0.0333 & 0.1049 & 0.0732 & 0.0439 & 0.1298 & 0.1306 & 0.6338 & 0.0913 & 0.0377 & 6.0393  \\
        CFGAN uZP               & 0.0230 & 0.1054 & 0.0228 & 0.0664 & 0.0579 & 0.0378 & 0.0843 & 0.1204 & 0.1238 & 0.0341 & 0.0169 & 4.5934  \\
        \bottomrule
    
\end{longtable}

\begin{table}[H]
    \centering
    \caption{Training and recommendation time comparison of baselines and CFGAN models for the Ciao dataset. CFGAN models use early-stopping.}
    \label{appendix:tab:ciao-reproducibility-scalability}
    \begin{tabular}{l|rrr}
    \toprule
    {} &                             Train Time & \begin{tabular}{@{}c@{}}Recommendation\\Time\end{tabular} & \begin{tabular}{@{}c@{}}Recommendation\\Throughput\end{tabular} \\
    \midrule
    Random                &                            0.00 [sec]  &                                        0.81 [sec]  &                                                793 \\
    TopPop                &                            0.00 [sec]  &                                        1.32 [sec]  &                                                486 \\
    \midrule
    UserKNN CF cosine     &                 0.04 $\pm$ 0.01 [sec]  &                             0.84 $\pm$ 0.25 [sec]  &                                               1095 \\
    UserKNN CF dice       &                 0.04 $\pm$ 0.01 [sec]  &                             0.98 $\pm$ 0.21 [sec]  &                                                507 \\
    UserKNN CF jaccard    &                 0.04 $\pm$ 0.01 [sec]  &                             1.01 $\pm$ 0.23 [sec]  &                                                492 \\
    UserKNN CF asymmetric &                 0.03 $\pm$ 0.01 [sec]  &                             1.00 $\pm$ 0.19 [sec]  &                                                732 \\
    UserKNN CF tversky    &                 0.04 $\pm$ 0.01 [sec]  &                             0.84 $\pm$ 0.11 [sec]  &                                                860 \\
    ItemKNN CF cosine     &                 0.10 $\pm$ 0.03 [sec]  &                             1.10 $\pm$ 0.43 [sec]  &                                                720 \\
    ItemKNN CF dice       &                 0.09 $\pm$ 0.04 [sec]  &                             1.02 $\pm$ 0.19 [sec]  &                                                551 \\
    ItemKNN CF jaccard    &                 0.08 $\pm$ 0.03 [sec]  &                             0.97 $\pm$ 0.15 [sec]  &                                                655 \\
    ItemKNN CF asymmetric &                 0.10 $\pm$ 0.04 [sec]  &                             1.13 $\pm$ 0.28 [sec]  &                                                413 \\
    ItemKNN CF tversky    &                 0.10 $\pm$ 0.04 [sec]  &                             1.00 $\pm$ 0.32 [sec]  &                                                707 \\
    RP3beta               &                 0.68 $\pm$ 0.25 [sec]  &                             0.96 $\pm$ 0.19 [sec]  &                                                835 \\
    PureSVD               &                 7.30 $\pm$ 6.22 [sec]  &                             1.69 $\pm$ 0.38 [sec]  &                                                300 \\
    SLIM ElasticNet       &                 5.45 $\pm$ 1.13 [sec]  &                             1.08 $\pm$ 0.01 [sec]  &                                                598 \\
    MF BPR                &               17.13 $\pm$ 12.22 [sec]  &                             0.99 $\pm$ 0.08 [sec]  &                                                688 \\
    EASE R                &                16.07 $\pm$ 3.69 [sec]  &                             1.18 $\pm$ 0.37 [sec]  &                                                750 \\
    \midrule
    CFGAN iZR             &  144.21 [sec] / 2.40 $\pm$ 0.82 [min]  &                             0.98 $\pm$ 0.31 [sec]  &                                                455 \\
    CFGAN iPM             &  166.97 [sec] / 2.78 $\pm$ 0.80 [min]  &                             0.95 $\pm$ 0.09 [sec]  &                                                581 \\
    CFGAN iZP             &  211.99 [sec] / 3.53 $\pm$ 1.02 [min]  &                             1.06 $\pm$ 0.31 [sec]  &                                                707 \\
    \midrule
    CFGAN uZR             &  112.04 [sec] / 1.87 $\pm$ 0.51 [min]  &                             1.02 $\pm$ 0.30 [sec]  &                                                617 \\
    CFGAN uPM             &  114.75 [sec] / 1.91 $\pm$ 0.44 [min]  &                             0.81 $\pm$ 0.09 [sec]  &                                                715 \\
    CFGAN uZP             &  158.07 [sec] / 2.63 $\pm$ 0.71 [min]  &                             0.95 $\pm$ 0.08 [sec]  &                                                694 \\
    \bottomrule
    \end{tabular}
\end{table}

\begin{table}[H]
    \centering
    \caption{Accuracy and beyond-accuracy metrics for tuned baselines and \gls{cfgan} on the ML100K dataset at
    recommendation list length of 20. \gls{cfgan} results are different than \autoref{appendix:tab:ml100k-concerns-at-20} due to the hyper-parameter tuning.}
    \label{appendix:tab:ml100k-reproducibility-@20}
    \begin{tabular}{l|ccccccc|ccccc}
        \toprule
        {}
        & PREC
        & REC
        & MAP
        & MRR
        & NDCG
        & F1
        & ARHR
        & Novelty
        & \begin{tabular}{@{}c@{}} Div. \\ MIL \end{tabular}
        & \begin{tabular}{@{}c@{}}               Cov. \\ Item \end{tabular}
        & \begin{tabular}{@{}c@{}}Div. \\ Gini \end{tabular}
        & \begin{tabular}{@{}c@{}}                        Div. \\ Shannon\end{tabular}       \\
        \midrule
        Random                  & 0.0153 & 0.0141 & 0.0044 & 0.0543 & 0.0185 & 0.0147 & 0.0600 & 0.1476 & 0.9878 & 1.0000 & 0.8303 & 10.6237 \\
        TopPop                  & 0.1560 & 0.1675 & 0.0971 & 0.4130 & 0.2132 & 0.1616 & 0.7151 & 0.0972 & 0.4526 & 0.0502 & 0.0207 & 5.3840 \\
        \midrule
        UserKNN CF cosine       & 0.2641 & 0.3239 & 0.2292 & 0.6581 & 0.3962 & 0.2910 & 1.3349 & 0.1031 & 0.8365 & 0.2432 & 0.0730 & 7.3278 \\
        UserKNN CF dice         & 0.2629 & 0.3218 & 0.2282 & 0.6514 & 0.3943 & 0.2894 & 1.3248 & 0.1029 & 0.8367 & 0.2335 & 0.0727 & 7.3199 \\
        UserKNN CF jaccard      & 0.2627 & 0.3227 & 0.2281 & 0.6545 & 0.3946 & 0.2896 & 1.3268 & 0.1029 & 0.8369 & 0.2353 & 0.0729 & 7.3238 \\
        UserKNN CF asymmetric   & 0.2582 & 0.3204 & 0.2220 & 0.6550 & 0.3891 & 0.2859 & 1.3059 & 0.1026 & 0.8237 & 0.2226 & 0.0663 & 7.1813 \\
        UserKNN CF tversky      & 0.2683 & 0.3257 & 0.2350 & 0.6609 & 0.4015 & 0.2942 & 1.3519 & 0.1035 & 0.8515 & 0.2662 & 0.0819 & 7.4997 \\
        ItemKNN CF cosine       & 0.2591 & 0.3189 & 0.2263 & 0.6469 & 0.3901 & 0.2859 & 1.3043 & 0.1031 & 0.8455 & 0.1984 & 0.0724 & 7.3031 \\
        ItemKNN CF dice         & 0.2454 & 0.3029 & 0.2124 & 0.6317 & 0.3719 & 0.2711 & 1.2497 & 0.1021 & 0.8165 & 0.1712 & 0.0613 & 7.0631 \\
        ItemKNN CF jaccard      & 0.2401 & 0.2943 & 0.2077 & 0.6334 & 0.3652 & 0.2644 & 1.2348 & 0.1014 & 0.7882 & 0.1525 & 0.0536 & 6.8701 \\
        ItemKNN CF asymmetric   & 0.2652 & 0.3266 & 0.2337 & 0.6496 & 0.3978 & 0.2927 & 1.3185 & 0.1062 & 0.8935 & 0.3479 & 0.1113 & 7.9516 \\
        ItemKNN CF tversky      & 0.2779 & 0.3372 & 0.2455 & 0.6615 & 0.4121 & 0.3047 & 1.3759 & 0.1036 & 0.8598 & 0.2329 & 0.0833 & 7.5266 \\
        RP3beta                 & 0.2603 & 0.3204 & 0.2286 & 0.6566 & 0.3928 & 0.2872 & 1.3184 & 0.1022 & 0.8174 & 0.1887 & 0.0629 & 7.1103 \\
        PureSVD                 & 0.2863 & 0.3451 & 0.2543 & 0.6667 & 0.4225 & 0.3130 & 1.4022 & 0.1053 & 0.8885 & 0.3037 & 0.1079 & 7.9044 \\
        SLIM ElasticNet         & 0.2915 & 0.3563 & 0.2683 & 0.6952 & 0.4375 & 0.3207 & 1.4585 & 0.1037 & 0.8597 & 0.2595 & 0.0860 & 7.5762 \\
        MF BPR                  & 0.2263 & 0.2829 & 0.1746 & 0.5620 & 0.3296 & 0.2515 & 1.0800 & 0.1017 & 0.7774 & 0.1682 & 0.0524 & 6.8514 \\
        EASE R                  & 0.2929 & 0.3530 & 0.2688 & 0.6909 & 0.4368 & 0.3202 & 1.4630 & 0.1042 & 0.8713 & 0.2783 & 0.0938 & 7.7031 \\
        \midrule
        CFGAN iZR               & 0.2415 & 0.3025 & 0.1923 & 0.5866 & 0.3546 & 0.2685 & 1.1513 & 0.1073 & 0.9017 & 0.3400 & 0.1235 & 8.0988 \\
        CFGAN iPM               & 0.2171 & 0.2350 & 0.1641 & 0.5247 & 0.3023 & 0.2257 & 1.0393 & 0.1037 & 0.8417 & 0.2390 & 0.0731 & 7.3341 \\
        CFGAN iZP               & 0.2757 & 0.3283 & 0.2394 & 0.6580 & 0.4041 & 0.2997 & 1.3595 & 0.1040 & 0.8572 & 0.2589 & 0.0848 & 7.5584 \\
        \midrule
        CFGAN uZR               & 0.2754 & 0.3387 & 0.2409 & 0.6500 & 0.4078 & 0.3038 & 1.3454 & 0.1060 & 0.8844 & 0.2371 & 0.1018 & 7.8011 \\
        CFGAN uPM               & 0.2326 & 0.3057 & 0.1864 & 0.5809 & 0.3498 & 0.2642 & 1.1201 & 0.1058 & 0.8699 & 0.2607 & 0.0951 & 7.7164 \\
        CFGAN uZP               & 0.2555 & 0.3163 & 0.2194 & 0.6399 & 0.3834 & 0.2827 & 1.2712 & 0.1055 & 0.8855 & 0.2529 & 0.1055 & 7.8565  \\
        \bottomrule
    \end{tabular}
\end{table}

\begin{table}[H]
    \centering
    \caption{Training and recommendation time comparison of baselines and CFGAN models for the ML100K dataset. CFGAN models use early-stopping.}
    \label{appendix:tab:ml100k-reproducibility-scalability}
    \begin{tabular}{l|rrr}
        \toprule
        {} & Train Time & \begin{tabular}{@{}c@{}}
                              Recommendation\\Time
        \end{tabular}                & \begin{tabular}{@{}c@{}}
                                           Recommendation\\Throughput
        \end{tabular}                \\
        \midrule
        Random     & 0.00 [sec]                & 1.24 [sec] & 759 \\
        TopPop       & 0.00 [sec]                & 1.71 [sec] & 551 \\
        \midrule
        UserKNN CF cosine    & 0.15 $\pm$ 0.05 [sec]                & 1.44 $\pm$ 0.18 [sec] & 634 \\
        UserKNN CF dice & 0.15 $\pm$ 0.05 [sec]                & 1.83 $\pm$ 0.43 [sec] & 617 \\
        UserKNN CF jaccard    & 0.15 $\pm$ 0.03 [sec]                & 1.78 $\pm$ 0.03 [sec] & 537 \\
        UserKNN CF asymmetric     & 0.16 $\pm$ 0.04 [sec]                & 1.81 $\pm$ 0.10 [sec] & 525 \\
        UserKNN CF tversky       & 0.15 $\pm$ 0.04 [sec]                & 1.45 $\pm$ 0.35 [sec] & 768 \\
        ItemKNN CF cosine    & 0.23 $\pm$ 0.06 [sec]                & 1.89 $\pm$ 0.12 [sec] & 475 \\
        ItemKNN CF dice & 0.21 $\pm$ 0.07 [sec]                & 1.95 $\pm$ 0.28 [sec] & 408 \\
        ItemKNN CF jaccard    & 0.21 $\pm$ 0.06 [sec]                & 1.69 $\pm$ 0.25 [sec] & 673 \\
        ItemKNN CF asymmetric               & 0.22 $\pm$ 0.05 [sec]                & 1.80 $\pm$ 0.17 [sec] & 557 \\
        ItemKNN CF tversky               & 0.22 $\pm$ 0.10 [sec]                & 1.77 $\pm$ 0.54 [sec] & 526 \\
        RP3beta       & 3.41 $\pm$ 2.22 [sec]               & 2.24 $\pm$ 0.83 [sec] & 328 \\
        PureSVD                & 4.32 $\pm$ 5.09 [sec]              & 2.30 $\pm$ 0.69 [sec] & 346 \\
        SLIM ElasticNet                & 13.32 $\pm$ 4.52 [sec]               & 2.24 $\pm$ 0.30 [sec] & 429 \\
        MF BPR             & 44.98 $\pm$ 51.80 [sec] & 1.78 $\pm$ 0.38 [sec] & 389 \\
        EASE R             & 14.33 $\pm$ 0.66 [sec] & 1.95 $\pm$ 0.17 [sec] & 498 \\
        \midrule
        CFGAN iZR             & 214.36 [sec] / 3.57 $\pm$ 0.86 [min] & 1.39 $\pm$ 0.06 [sec] & 699 \\
        CFGAN iPM             & 198.47 [sec] / 3.31 $\pm$ 0.74 [min] & 2.10 $\pm$ 0.46 [sec] & 531 \\
        CFGAN iZP             & 339.04 [sec] / 5.65 $\pm$ 1.61 [min] & 1.73 $\pm$ 0.16 [sec] & 564 \\
        CFGAN uZR             & 182.70 [sec] / 3.04 $\pm$ 1.01 [min] & 1.55 $\pm$ 0.03 [sec] & 617 \\
        CFGAN uPM             & 194.96 [sec] / 3.25 $\pm$ 0.85 [min] & 1.66 $\pm$ 0.24 [sec] & 612 \\
        CFGAN uZP             & 258.30 [sec] / 4.31 $\pm$ 0.86 [min] & 1.82 $\pm$ 0.51 [sec] & 577 \\
        \bottomrule
    \end{tabular}

\end{table}

\begin{table}[H]
    \centering
    \caption{Accuracy and beyond-accuracy metrics for tuned baselines and \gls{cfgan} on the ML1M dataset at
    recommendation list length of 20. \gls{cfgan} results are different than \autoref{appendix:tab:ml1m-concerns-at-20} due to the hyper-parameter tuning.}
    \label{appendix:tab:ml1m-reproducibility-@20}
    \begin{tabular}{l|ccccccc|ccccc}
        \toprule
        {}
        & PREC
        & REC
        & MAP
        & MRR
        & NDCG
        & F1
        & ARHR
        & Novelty
        & \begin{tabular}{@{}c@{}} Div. \\ MIL \end{tabular}
        & \begin{tabular}{@{}c@{}}               Cov. \\ Item \end{tabular}
        & \begin{tabular}{@{}c@{}}Div. \\ Gini \end{tabular}
        & \begin{tabular}{@{}c@{}}                        Div. \\ Shannon\end{tabular}       \\
        \midrule
        Random                  & 0.0099 & 0.0056 & 0.0024 & 0.0326 & 0.0108 & 0.0072 & 0.0364 & 0.0732 & 0.9946 & 1.0000 & 0.8977 & 11.8223 \\
        TopPop                  & 0.1552 & 0.1146 & 0.0917 & 0.3852 & 0.1938 & 0.1319 & 0.7065 & 0.0473 & 0.4529 & 0.0299 & 0.0095 & 5.4298 \\
        \midrule
        UserKNN CF cosine       & 0.2883 & 0.2525 & 0.2265 & 0.6505 & 0.3855 & 0.2693 & 1.3785 & 0.0520 & 0.9110 & 0.4329 & 0.0828 & 8.6076 \\
        UserKNN CF dice         & 0.2877 & 0.2551 & 0.2252 & 0.6556 & 0.3863 & 0.2704 & 1.3802 & 0.0512 & 0.8890 & 0.3194 & 0.0644 & 8.2557 \\
        UserKNN CF jaccard      & 0.2893 & 0.2565 & 0.2273 & 0.6571 & 0.3884 & 0.2719 & 1.3866 & 0.0514 & 0.8971 & 0.3528 & 0.0702 & 8.3766 \\
        UserKNN CF asymmetric   & 0.2891 & 0.2570 & 0.2273 & 0.6595 & 0.3888 & 0.2721 & 1.3894 & 0.0513 & 0.8921 & 0.3286 & 0.0655 & 8.2848 \\
        UserKNN CF tversky      & 0.2890 & 0.2561 & 0.2272 & 0.6569 & 0.3881 & 0.2715 & 1.3850 & 0.0513 & 0.8948 & 0.3408 & 0.0683 & 8.3392 \\
        ItemKNN CF cosine       & 0.2785 & 0.2392 & 0.2138 & 0.6279 & 0.3688 & 0.2574 & 1.3205 & 0.0515 & 0.9018 & 0.3197 & 0.0687 & 8.3666 \\
        ItemKNN CF dice         & 0.2566 & 0.2110 & 0.1913 & 0.5940 & 0.3376 & 0.2316 & 1.2253 & 0.0503 & 0.8519 & 0.2531 & 0.0455 & 7.7630 \\
        ItemKNN CF jaccard      & 0.2556 & 0.2108 & 0.1908 & 0.5925 & 0.3369 & 0.2310 & 1.2226 & 0.0503 & 0.8453 & 0.2596 & 0.0445 & 7.7211 \\
        ItemKNN CF asymmetric   & 0.2600 & 0.2196 & 0.1985 & 0.6254 & 0.3490 & 0.2381 & 1.2744 & 0.0497 & 0.8148 & 0.2097 & 0.0362 & 7.4341 \\
        ItemKNN CF tversky      & 0.2657 & 0.2180 & 0.2005 & 0.6147 & 0.3496 & 0.2395 & 1.2710 & 0.0513 & 0.9005 & 0.1814 & 0.0560 & 8.1173 \\
        RP3beta                 & 0.2758 & 0.2385 & 0.2146 & 0.6425 & 0.3700 & 0.2558 & 1.3346 & 0.0506 & 0.8565 & 0.3427 & 0.0528 & 7.9254 \\
        PureSVD                 & 0.2913 & 0.2421 & 0.2234 & 0.6333 & 0.3783 & 0.2644 & 1.3555 & 0.0516 & 0.9142 & 0.2439 & 0.0712 & 8.4463 \\
        SLIM ElasticNet         & 0.3119 & 0.2695 & 0.2508 & 0.6724 & 0.4123 & 0.2892 & 1.4658 & 0.0514 & 0.8984 & 0.3153 & 0.0696 & 8.3690 \\
        MF BPR                  & 0.2485 & 0.2103 & 0.1762 & 0.5753 & 0.3242 & 0.2278 & 1.1594 & 0.0512 & 0.8855 & 0.3126 & 0.0631 & 8.2195 \\
        EASE R                  & 0.3171 & 0.2763 & 0.2560 & 0.6795 & 0.4192 & 0.2953 & 1.4853 & 0.0518 & 0.9146 & 0.3338 & 0.0803 & 8.5897 \\
        \midrule
        CFGAN iZR               & 0.2862 & 0.2547 & 0.2146 & 0.6312 & 0.3770 & 0.2696 & 1.3288 & 0.0542 & 0.9583 & 0.4123 & 0.1459 & 9.4737 \\
        CFGAN iPM               & 0.2505 & 0.1950 & 0.1734 & 0.5454 & 0.3138 & 0.2193 & 1.1354 & 0.0523 & 0.9218 & 0.3669 & 0.0901 & 8.7458 \\
        CFGAN iZP               & 0.2407 & 0.1742 & 0.1661 & 0.5230 & 0.2972 & 0.2021 & 1.0929 & 0.0530 & 0.9256 & 0.4894 & 0.0901 & 8.7580 \\
        \midrule
        CFGAN uZR               & 0.2955 & 0.2473 & 0.2241 & 0.6222 & 0.3799 & 0.2692 & 1.3541 & 0.0523 & 0.9205 & 0.2167 & 0.0837 & 8.6304 \\
        CFGAN uPM               & 0.2367 & 0.1928 & 0.1629 & 0.5513 & 0.3054 & 0.2125 & 1.1064 & 0.0516 & 0.8962 & 0.1782 & 0.0550 & 8.0858 \\
        CFGAN uZP               & 0.2764 & 0.2342 & 0.2074 & 0.6208 & 0.3620 & 0.2536 & 1.3010 & 0.0513 & 0.9062 & 0.1833 & 0.0617 & 8.2408  \\
        \bottomrule
    \end{tabular}
\end{table}

\begin{table}[H]
    \centering
    \caption{Training and recommendation time comparison of baselines and CFGAN models for the ML1M dataset. CFGAN models use early-stopping.}
    \label{appendix:tab:ml1m-reproducibility-scalability}
    \begin{tabular}{l|rrr}
        \toprule
        {} & Train Time & \begin{tabular}{@{}c@{}}
                              Recommendation\\Time
        \end{tabular}                & \begin{tabular}{@{}c@{}}
                                           Recommendation\\Throughput
        \end{tabular}                \\
        \midrule
        Random     & 0.02 [sec]                   & 12.52 [sec] & 482 \\
        TopPop       & 0.02 [sec]                   & 13.97 [sec] & 432 \\
        \midrule
        UserKNN CF cosine    & 5.93 $\pm$ 2.09 [sec]                   & 17.86 $\pm$ 4.00 [sec] & 389 \\
        UserKNN CF dice & 4.01 $\pm$ 0.60 [sec]                   & 12.78 $\pm$ 0.66 [sec] & 483 \\
        UserKNN CF jaccard    & 5.44 $\pm$ 1.67 [sec]                   & 19.57 $\pm$ 5.00 [sec] & 354 \\
        UserKNN CF asymmetric     & 4.18 $\pm$ 0.55 [sec]                   & 11.99 $\pm$ 0.21 [sec] & 497 \\
        UserKNN CF tversky       & 4.84 $\pm$ 0.90 [sec]                   & 13.19 $\pm$ 1.05 [sec] & 424 \\
        ItemKNN CF cosine    & 2.75 $\pm$ 1.00 [sec]                   & 16.33 $\pm$ 4.50 [sec] & 413 \\
        ItemKNN CF dice & 1.85 $\pm$ 0.43 [sec]                   & 13.24 $\pm$ 0.29 [sec] & 463 \\
        ItemKNN CF jaccard    & 2.76 $\pm$ 0.98 [sec]                   & 20.45 $\pm$ 2.44 [sec] & 269 \\
        ItemKNN CF asymmetric               & 2.10 $\pm$ 0.56 [sec]                   & 14.09 $\pm$ 0.69 [sec] & 453 \\
        ItemKNN CF tversky               & 2.00 $\pm$ 0.37 [sec]                 & 12.63 $\pm$ 0.56 [sec] & 452 \\
        RP3beta       & 9.69 $\pm$ 4.41 [sec]    & 18.74 $\pm$ 4.75 [sec] & 290 \\
        PureSVD                & 10.27 $\pm$ 10.68 [sec]    & 12.01 $\pm$ 2.38 [sec] & 545 \\
        SLIM ElasticNet                & 204.75 [sec] / 3.41 $\pm$ 2.27 [min]                  & 16.44 $\pm$ 5.32 [sec] & 467 \\
        MF BPR             & 497.53 [sec] / 8.29 $\pm$ 8.04 [min]  & 11.03 $\pm$ 1.80 [sec]  & 534 \\
        EASE R             & 24.26 $\pm$ 3.32 [sec]  & 14.00 $\pm$ 1.43 [sec]  & 402 \\
        \midrule
        CFGAN iZR             & 1554.32 [sec] / 25.91 $\pm$ 7.34 [min] & 9.96 $\pm$ 0.79 [sec]  & 691 \\
        CFGAN iPM             & 1751.51 [sec] / 29.19 $\pm$ 8.93 [min]  & 8.46 $\pm$ 1.11 [sec]  & 889 \\
        CFGAN iZP             & 2100.12 [sec] / 35.00 $\pm$ 10.09 [min]  & 8.80 $\pm$ 2.21 [sec]  & 896 \\
        \midrule
        CFGAN uZR             & 1602.18 [sec] / 26.70 $\pm$ 7.26 [min]  & 9.16 $\pm$ 0.79 [sec]  & 656 \\
        CFGAN uPM             & 1637.66 [sec] / 27.29 $\pm$ 6.76 [min]  & 9.52 $\pm$ 1.05 [sec]  & 687 \\
        CFGAN uZP             & 2221.04 [sec] / 37.02 $\pm$ 9.31 [min]  & 9.13 $\pm$ 2.55 [sec]  & 906 \\
        \bottomrule
    \end{tabular}
\end{table}

\end{landscape}

%
%
\bibliographystyle{splncs04}
\bibliography{main}

\begin{thebibliography}{10}
\providecommand{\url}[1]{\texttt{#1}}
\providecommand{\urlprefix}{URL }
\providecommand{\doi}[1]{https://doi.org/#1}

\bibitem{DBLP:journals/tkde/AdomaviciusK12/beyond-accuracy/diversity-gini}
Adomavicius, G., Kwon, Y.: Improving aggregate recommendation diversity using
  ranking-based techniques. {IEEE} Trans. Knowl. Data Eng.  \textbf{24}(5),
  896--911 (2012). \doi{10.1109/TKDE.2011.15}

\bibitem{improvements-that-dont-add-up-ad-hoc-retrieval-results-since-1998}
Armstrong, T.G., Moffat, A., Webber, W., Zobel, J.: Improvements that don't add
  up: ad-hoc retrieval results since 1998. In: Proceedings of the 18th {ACM}
  Conference on Information and Knowledge Management, {CIKM} 2009, Hong Kong,
  China, November 2-6, 2009. pp. 601--610. {ACM} (2009).
  \doi{10.1145/1645953.1646031}

\bibitem{gan-evaluation}
Borji, A.: Pros and cons of {GAN} evaluation measures. Comput. Vis. Image
  Underst.  \textbf{179},  41--65 (2019). \doi{10.1016/j.cviu.2018.10.009}

\bibitem{cfgan}
Chae, D., Kang, J., Kim, S., Lee, J.: {CFGAN:} {A} generic collaborative
  filtering framework based on generative adversarial networks. In: Proceedings
  of the 27th {ACM} International Conference on Information and Knowledge
  Management, {CIKM} 2018, Torino, Italy, October 22-26, 2018. pp. 137--146.
  {ACM} (2018). \doi{10.1145/3269206.3271743}

\bibitem{tagrec-gan}
Chen, H., Wang, S., Jiang, N., Li, Z., Yan, N., Shi, L.: Trust-aware generative
  adversarial network with recurrent neural network for recommender systems.
  Int. J. Intell. Syst.  \textbf{36}(2),  778--795 (2021).
  \doi{10.1002/int.22320}

\bibitem{rp3-beta-blockbusters-and-wallflowers-accurate-diverse-and-scalable-recommendations-with-random-walks}
Christoffel, F., Paudel, B., Newell, C., Bernstein, A.: Blockbusters and
  wallflowers: Accurate, diverse, and scalable recommendations with random
  walks. In: Proceedings of the 9th {ACM} Conference on Recommender Systems,
  RecSys 2015, Vienna, Austria, September 16-20, 2015. pp. 163--170. {ACM}
  (2015). \doi{10.1145/2792838.2800180}

\bibitem{performace-of-recommender-algorithms-on-top-n-recommendation-tasks}
Cremonesi, P., Koren, Y., Turrin, R.: Performance of recommender algorithms on
  top-n recommendation tasks. In: Proceedings of the 2010 {ACM} Conference on
  Recommender Systems, RecSys 2010, Barcelona, Spain, September 26-30, 2010.
  pp. 39--46. {ACM} (2010). \doi{10.1145/1864708.1864721}

\bibitem{gans-an-overview}
Creswell, A., White, T., Dumoulin, V., Arulkumaran, K., Sengupta, B., Bharath,
  A.A.: Generative adversarial networks: An overview. {IEEE} Signal Process.
  Mag.  \textbf{35}(1),  53--65 (2018). \doi{10.1109/MSP.2017.2765202}

\bibitem{seed-2-effects-of-random-seeds-on-the-accuracy-of-convolutional-neural-networks}
Fellicious, C., Wei{\ss}gerber, T., Granitzer, M.: Effects of random seeds on
  the accuracy of convolutional neural networks. In: Machine Learning,
  Optimization, and Data Science - 6th International Conference, {LOD} 2020,
  Siena, Italy, July 19-23, 2020, Revised Selected Papers, Part {II}. Lecture
  Notes in Computer Science, vol. 12566, pp. 93--102. Springer (2020).
  \doi{10.1007/978-3-030-64580-9\_8}

\bibitem{a-troubling-analysis-of-reproducbility-and-progress-in-recsys-research}
{Ferrari Dacrema}, M., Boglio, S., Cremonesi, P., Jannach, D.: A troubling
  analysis of reproducibility and progress in recommender systems research.
  {ACM} Trans. Inf. Syst.  \textbf{39}(2),  20:1--20:49 (2021).
  \doi{10.1145/3434185}

\bibitem{are-we-really-making-much-progress-a-worrying-analysis-of-recent-neural-recommendation-approaches}
{Ferrari Dacrema}, M., Cremonesi, P., Jannach, D.: Are we really making much
  progress? {A} worrying analysis of recent neural recommendation approaches.
  In: Proceedings of the 13th {ACM} Conference on Recommender Systems, RecSys
  2019, Copenhagen, Denmark, September 16-20, 2019. pp. 101--109. {ACM} (2019).
  \doi{10.1145/3298689.3347058}

\bibitem{critically-examining-the-claimed-value-of-convolutions-over-user-item-embedding-maps-for-recommender-systems}
{Ferrari Dacrema}, M., Parroni, F., Cremonesi, P., Jannach, D.: Critically
  examining the claimed value of convolutions over user-item embedding maps for
  recommender systems. In: {CIKM} '20: The 29th {ACM} International Conference
  on Information and Knowledge Management, Virtual Event, Ireland, October
  19-23, 2020. pp. 355--363. {ACM} (2020). \doi{10.1145/3340531.3411901}

\bibitem{gan}
Goodfellow, I.J., Pouget{-}Abadie, J., Mirza, M., Xu, B., Warde{-}Farley, D.,
  Ozair, S., Courville, A.C., Bengio, Y.: Generative adversarial nets. In:
  Advances in Neural Information Processing Systems 27: Annual Conference on
  Neural Information Processing Systems 2014, December 8-13 2014, Montreal,
  Quebec, Canada. pp. 2672--2680 (2014),
  \url{https://proceedings.neurips.cc/paper/2014/hash/5ca3e9b122f61f8f06494c97b1afccf3-Abstract.html}

\bibitem{gan-communications-acm}
Goodfellow, I.J., Pouget{-}Abadie, J., Mirza, M., Xu, B., Warde{-}Farley, D.,
  Ozair, S., Courville, A.C., Bengio, Y.: Generative adversarial networks.
  Commun. {ACM}  \textbf{63}(11),  139--144 (2020). \doi{10.1145/3422622}

\bibitem{movielens}
Harper, F.M., Konstan, J.A.: The {MovieLens} datasets: History and context.
  {ACM} Trans. Interact. Intell. Syst.  \textbf{5}(4),  19:1--19:19 (2016).
  \doi{10.1145/2827872}

\bibitem{image-to-image-conditional-gan}
Isola, P., Zhu, J., Zhou, T., Efros, A.A.: Image-to-image translation with
  conditional adversarial networks. In: 2017 {IEEE} Conference on Computer
  Vision and Pattern Recognition, {CVPR} 2017, Honolulu, HI, USA, July 21-26,
  2017. pp. 5967--5976. {IEEE} Computer Society (2017).
  \doi{10.1109/CVPR.2017.632}

\bibitem{stylegan}
Karras, T., Laine, S., Aila, T.: A style-based generator architecture for
  generative adversarial networks. In: {IEEE} Conference on Computer Vision and
  Pattern Recognition, {CVPR} 2019, Long Beach, CA, USA, June 16-20, 2019. pp.
  4401--4410. Computer Vision Foundation / {IEEE} (2019).
  \doi{10.1109/CVPR.2019.00453}

\bibitem{stylegan2}
Karras, T., Laine, S., Aittala, M., Hellsten, J., Lehtinen, J., Aila, T.:
  Analyzing and improving the image quality of {StyleGAN}. In: 2020 {IEEE/CVF}
  Conference on Computer Vision and Pattern Recognition, {CVPR} 2020, Seattle,
  WA, USA, June 13-19, 2020. pp. 8107--8116. Computer Vision Foundation /
  {IEEE} (2020). \doi{10.1109/CVPR42600.2020.00813}

\bibitem{adam-a-method-for-stochastic-optimization}
Kingma, D.P., Ba, J.: Adam: {A} method for stochastic optimization. In: 3rd
  International Conference on Learning Representations, {ICLR} 2015, San Diego,
  CA, USA, May 7-9, 2015, Conference Track Proceedings (2015)

\bibitem{DBLP:journals/ftml/KingmaW19/variational-autoencoder}
Kingma, D.P., Welling, M.: An introduction to variational autoencoders. Found.
  Trends Mach. Learn.  \textbf{12}(4),  307--392 (2019).
  \doi{10.1561/2200000056}

\bibitem{DBLP:journals/sigir/Lin18}
Lin, J.: The neural hype and comparisons against weak baselines. SIGIR Forum
  \textbf{52}(2),  40–51 (jan 2019). \doi{10.1145/3308774.3308781}

\bibitem{DBLP:journals/sigir/Lin19}
Lin, J.: The neural hype, justified! a recantation. SIGIR Forum
  \textbf{53}(2),  88–93 (mar 2021). \doi{10.1145/3458553.3458563}

\bibitem{troubling-trends-in-machine-learning-scholarship}
Lipton, Z.C., Steinhardt, J.: Troubling trends in machine learning scholarship.
  {ACM} Queue  \textbf{17}(1), ~80 (2019). \doi{10.1145/3317287.3328534}

\bibitem{evaluation-of-session-bases-recommendation-algorithms}
Ludewig, M., Jannach, D.: Evaluation of session-based recommendation
  algorithms. User Model. User Adapt. Interact.  \textbf{28}(4-5),  331--390
  (2018). \doi{10.1007/s11257-018-9209-6}

\bibitem{seed-1-on-model-stability-as-a-function-of-random-seed}
Madhyastha, P., Jain, R.: On model stability as a function of random seed. In:
  Proceedings of the 23rd Conference on Computational Natural Language
  Learning, CoNLL 2019, Hong Kong, China, November 3-4, 2019. pp. 929--939.
  Association for Computational Linguistics (2019). \doi{10.18653/v1/K19-1087}

\bibitem{conditional-gan}
Mirza, M., Osindero, S.: Conditional generative adversarial nets. CoRR
  \textbf{abs/1411.1784} (2014), \url{http://arxiv.org/abs/1411.1784}

\bibitem{a-unifying-view-on-dataset-shift-in-classification}
Moreno{-}Torres, J.G., Raeder, T., Ala{\'{\i}}z{-}Rodr{\'{\i}}guez, R., Chawla,
  N.V., Herrera, F.: A unifying view on dataset shift in classification.
  Pattern Recognit.  \textbf{45}(1),  521--530 (2012).
  \doi{10.1016/j.patcog.2011.06.019}

\bibitem{slim}
Ning, X., Karypis, G.: {SLIM:} sparse linear methods for top-n recommender
  systems. In: 11th {IEEE} International Conference on Data Mining, {ICDM}
  2011, Vancouver, BC, Canada, December 11-14, 2011. pp. 497--506. {IEEE}
  Computer Society (2011). \doi{10.1109/ICDM.2011.134}

\bibitem{supplemental_material}
Pérez~Maurera, F.B., Ferrari~Dacrema, M., Cremonesi, P.: {An Evaluation of
  Generative Adversarial Networks for Collaborative Filtering - Supplemental
  Material} (Jan 2022). \doi{10.5281/zenodo.5879345}

\bibitem{dataset-shift-in-machine-learning}
Quionero-Candela, J., Sugiyama, M., Schwaighofer, A., Lawrence, N.D.: Dataset
  Shift in Machine Learning. The MIT Press (2009)

\bibitem{bpr-bayesian-personalized-ranking-from-implicit-feedback}
Rendle, S., Freudenthaler, C., Gantner, Z., Schmidt{-}Thieme, L.: {BPR:}
  bayesian personalized ranking from implicit feedback. In: {UAI} 2009,
  Proceedings of the Twenty-Fifth Conference on Uncertainty in Artificial
  Intelligence, Montreal, QC, Canada, June 18-21, 2009. pp. 452--461. {AUAI}
  Press (2009)

\bibitem{ease-r-embarrassingly-shallow-autoencoders-for-sparse-data}
Steck, H.: Embarrassingly shallow autoencoders for sparse data. In: The World
  Wide Web Conference, {WWW} 2019, San Francisco, CA, USA, May 13-17, 2019. pp.
  3251--3257. {ACM} (2019). \doi{10.1145/3308558.3313710}

\bibitem{ciao-dataset}
Tang, J., Gao, H., Liu, H.: {mTrust:} discerning multi-faceted trust in a
  connected world. In: Proceedings of the Fifth International Conference on Web
  Search and Web Data Mining, {WSDM} 2012, Seattle, WA, USA, February 8-12,
  2012. pp. 93--102. {ACM} (2012). \doi{10.1145/2124295.2124309}

\bibitem{irgan}
Wang, J., Yu, L., Zhang, W., Gong, Y., Xu, Y., Wang, B., Zhang, P., Zhang, D.:
  {IRGAN:} {A} minimax game for unifying generative and discriminative
  information retrieval models. In: Proceedings of the 40th International {ACM}
  {SIGIR} Conference on Research and Development in Information Retrieval,
  Shinjuku, Tokyo, Japan, August 7-11, 2017. pp. 515--524. {ACM} (2017).
  \doi{10.1145/3077136.3080786}

\bibitem{crgan-adversarial-preference-learning-with-pairwise-comparisons}
Wang, Z., Xu, Q., Ma, K., Jiang, Y., Cao, X., Huang, Q.: Adversarial preference
  learning with pairwise comparisons. In: Proceedings of the 27th {ACM}
  International Conference on Multimedia, {MM} 2019, Nice, France, October
  21-25, 2019. pp. 656--664. {ACM} (2019). \doi{10.1145/3343031.3350919}

\bibitem{mtpr-gan}
Xia, B., Bai, Y., Yin, J., Li, Q., Xu, L.: {MTPR}: A multi-task learning based
  poi recommendation considering temporal check-ins and geographical locations.
  Applied Sciences  \textbf{10}(19) (2020). \doi{10.3390/app10196664}

\bibitem{cfgan-service-recommendations}
Xie, F., Li, S., Chen, L., Xu, Y., Zheng, Z.: Generative adversarial network
  based service recommendation in heterogeneous information networks. In: 2019
  {IEEE} International Conference on Web Services, {ICWS} 2019, Milan, Italy,
  July 8-13, 2019. pp. 265--272. {IEEE} (2019). \doi{10.1109/ICWS.2019.00053}

\bibitem{DBLP:conf/sigir/YangLYL19}
Yang, W., Lu, K., Yang, P., Lin, J.: Critically examining the "neural hype":
  Weak baselines and the additivity of effectiveness gains from neural ranking
  models. In: Proceedings of the 42nd International {ACM} {SIGIR} Conference on
  Research and Development in Information Retrieval, {SIGIR} 2019, Paris,
  France, July 21-25, 2019. pp. 1129--1132. {ACM} (2019).
  \doi{10.1145/3331184.3331340}

\bibitem{beyond-accuracy/diversity-mean-inter-list}
Zhou, T., Kuscsik, Z., Liu, J.G., Medo, M., Wakeling, J.R., Zhang, Y.C.:
  Solving the apparent diversity-accuracy dilemma of recommender systems.
  Proceedings of the National Academy of Sciences  \textbf{107}(10),
  4511--4515 (2010). \doi{10.1073/pnas.1000488107}

\end{thebibliography}

\end{document}